\documentclass[fleqn,usenatbib]{mnras}

\usepackage{newtxtext,newtxmath}

\usepackage[T1]{fontenc}

\DeclareRobustCommand{\VAN}[3]{#2}
\let\VANthebibliography\thebibliography
\def\thebibliography{\DeclareRobustCommand{\VAN}[3]{##3}\VANthebibliography}

\usepackage{graphicx}	
\usepackage{amsmath}	
\usepackage{bm}
\usepackage[version=3]{mhchem}
\usepackage{booktabs}
\usepackage{floatrow}
\usepackage{subfig}
\usepackage{wrapfig}
\usepackage{capt-of}
\usepackage{xcolor}
\usepackage{placeins}
\usepackage{csquotes}
\usepackage{url}

\title[Cool Gaseous Exoplanets]{Cool Gaseous Exoplanets:  surveying the new frontier with Twinkle}

\author[L. Booth et al.]{
Luke Booth,$^{1}$\thanks{boothls@cardiff.ac.uk}
Subhajit Sarkar,$^{1}$
Matt Griffin,$^{1}$
and Billy Edwards,$^{2}$
\\
$^{1}$Cardiff Hub for Astrophysics Research and Technology (CHART), School of Physics and Astronomy, Cardiff University, 5 The Parade, CF24 3AA, United Kingdom\\
$^{2}$SRON, Netherlands Institute for Space Research, Niels Bohrweg 4, NL-2333 CA, Leiden, The Netherlands}

\date{Accepted XXX. Received YYY; in original form ZZZ}

\pubyear{2024}

\begin{document}
\label{firstpage}
\pagerange{\pageref{firstpage}--\pageref{lastpage}}
\maketitle

\begin{abstract}
Cool gaseous exoplanets ($1.75\ R_\oplus <  R_\text{p} < 3\ R_\text{J}$, $200$ K $<T_\text{eq} < 1000$~K) are an as-yet understudied population, with great potential to expand our understanding of planetary atmospheres and formation mechanisms. In this paper, we outline the basis for a homogeneous survey of cool gaseous planets with \textit{Twinkle}, a 0.45-m diameter space telescope with simultaneous spectral coverage from 0.5-4.5~$\mu$m, set to launch in 2025. 
We find that \textit{Twinkle} has the potential to characterise the atmospheres of 36 known cool gaseous exoplanets (11~sub-Neptunian, 11~Neptunian, 14~Jovian) at an SNR $\geq$ 5 during its 3-year primary mission, with the capability of detecting most major molecules predicted by equilibrium chemistry to > $5\sigma$ significance.  We find that an injected mass-metallicity trend is well-recovered, demonstrating \textit{Twinkle}'s ability to elucidate this fundamental relationship into cool regime. We also find \textit{Twinkle} will be able to detect cloud layers at 3$\sigma$ or greater in all cool gaseous planets for clouds at $\leq$ 10 Pa pressure level, but will be insensitive to clouds deeper than $10^4$ Pa in all cases. With these results we demonstrate the capability of the \textit{Twinkle} mission to greatly expand the current knowledge of cool gaseous planets, enabling key insights and constraints to be obtained for this poorly-charted region of exoplanet parameter space.
\end{abstract}



\section{Introduction}
Over the past three decades, the field of exoplanet science has progressed rapidly, from the first detections in the 1990s \citep{2012Wolszczan_PSR1257, 1995Mayor_and_Queloz_Nature} and the first atmospheric spectrum in 2002 \citep{FirstSpectra_Charbonneau2002}, to the revolution in exoplanet demographics resulting from \textit{Kepler} \citep{Kepler_Advances_Lissauer_2014, RadiusValley_Fulton_2017} and over a decade of transmission spectra from the \textit{Hubble Space Telescope} (HST) \citep{2022arXiv_HSTTrends_70}. Most recently, the \textit{James Webb Space Telescope} (JWST) is now returning high precision transmission spectra, resulting in new discoveries such as the first detection of sulphur-bearing species in an atmosphere  \citep{2023_WASP39b_(SO2)_JWST/NIRSpec-ERS_Nature1, 2023_WASP39b_(SO2)_JWST/NIRSpec-ERS_Nature_2, 2023_WASP39b_(SO2_photochemistry)_JWST/NIRSpec_Nature} and the first detection of carbon-bearing molecules in a habitable zone planet \citep{K2-18b_CH4_DMS_Madhusudhan2023}. Today over 5500\footnotemark[1]\footnotetext[1]{NASA Archive Planetary Systems Composite Table \url{https://exoplanetarchive.ipac.caltech.edu/cgi-bin/TblView/nph-tblView?app=ExoTbls&config=PSCompPars} (27/09/2023)} confirmed exoplanets are known, the majority of which have been detected by transit photometry using large dedicated space-based surveys such as \textit{Kepler}, K2, \textit{CHEOPS} and \textit{TESS} or ground-based surveys such as WASP, HATNet and NGTS. Such transiting planets provide potential targets for atmospheric characterization through transmission and/or eclipse spectroscopy. Using these techniques, in the next decade, the upcoming Ariel and \textit{Twinkle} missions will perform the first dedicated population-level surveys of exoplanet atmospheres \citep{ARIEL_Tinetti_2018, Twinkle_Edwards_2019}.   

The currently known transiting exoplanet population is highly diverse in both radius and temperature, containing planets with radii that vary from less than that of Mercury to several times the size of Jupiter and equilibrium temperatures, $T_\text{eq}$, that span the range from less than $200$ to over $4000$~K. Selection effects of the two most prolific detection methods (transit and radial velocity) bias the currently discovered planetary population towards shorter period planets.  Super-Earths and sub-Neptunes are the most frequent type of planet, often found in closely packed multi-planet systems, e.g. the TRAPPIST-1 \citep{TRAPPIST-1Discovery_Gillon&Triaud_2017}, Kepler-296 \citep{Kepler-296Discovery_Barclay_2015}, Kepler-32 \citep{Kepler-32Discovery_Swift_2013} and K2-384 \citep{K2-384Discovery_Christiansen_2022} systems.  
A large population of "hot Jupiters" is also known, their large size, short periods and high transit probabilities positively biasing their transit detectability.  Notably, some planet populations appear more sparse.
These include the "hot Neptune desert" \citep{HotNeptuneDesert_Szabo&Kiss_2011, HotNeptuneDesert_Mazeh_2016, NeptunianDesert_LTT9779b_Edwards2023} as well as colder gas giants. \cite{Wittenmyer2020}, relying on long-duration radial velocity data, determined that the occurrence rates of giant planets around solar-type stars was
fairly constant below 300~days and was increased at longer periods. The occurrence rates of hot Jupiters and 
temperate gas giants would thus be similar ($\sim$~1\%) but very cold giants, analogous to Jupiter or Saturn, are more frequent ($\sim$~7\%). Despite this, in practice there is a dearth of known transiting gas giants at cooler temperatures, with about half as many transiting giants having $T_\text{eq}$ less than 1000~K as those with $T_\text{eq}$ greater than 1000~K (around all types of star) (\autoref{tab: Transiting Planets by Characterisation}). 

\newpage
To date only a small fraction of all known transiting exoplanets, $\sim$ 180 planets\footnotemark[2]\footnotetext[2]{ExoAtmospheres community database (13/11/2023) \url{http://research.iac.es/proyecto/exoatmospheres/index.php}}, have had their atmospheres characterised through a combination of transmission, emission, cross-correlation and direct imaging spectroscopy, with specific molecular detections reported in about half these cases\footnotemark[3]\footnotetext[3]{Defined as where one or more molecules have definitive detections reported in the ExoAtmospheres community database, but excluding cases where only upper limits are given.}.  
The most successful method applied to date has been transmission spectroscopy. This technique is most sensitive to high scale height atmospheres and large planetary radii which tend to augment spectral feature amplitudes. 
The amplitude of spectral features in transmission $A_\text{p}$ can be approximated by:
\begin{equation}
\label{eqn: Spectral Amplitude}
    A_\text{p} = \frac{2 R_\text{p} \cdot nH}{R_\text{s}^2} 
\end{equation}
where $R_\text{p}$ is the radius of the planet, $R_\text{s}$ is the radius of the host star and $H$ is the pressure scale height. $n$ is commonly taken to have a value of 5, and gives the number of scales heights of a typical spectral features. 
The scale height is given by:
\begin{equation}
    \label{eqn: Scale Height}
    H = \frac{k_\text{B} T_\text{eq}}{\mu g}
\end{equation}
where $\mu$ is the mean molecular weight of the atmosphere, commonly taken to be $\sim 2.3$ for hydrogen/helium-dominated  (H$_2$-He) atmospheres, $g$ is the surface gravity of the planet, $k_\text{B}$ is Boltzmann's constant and $T_\text{eq}$ is the equilibrium temperature of the planet.  As well as making the best targets for transmission spectroscopy, large, very hot planets also make better targets for day-side emission spectroscopy due to their higher thermal flux. As a result, currently almost two thirds of all planets that have had their atmospheres analysed have $T_\text{eq}$ > $1000$ K, with this population accounting for $\sim$70\% of planets that have molecular detections\footnotemark[3]. 

Early searches for trends in exoplanet atmospheres \citep{H2OTrend_Sing2016, HSTPopulationStudy_Tsiaras2018, 2022arXiv_HSTTrends_70} have also typically been dominated by hot Jupiters and warm Neptunes. Trends reported include 
temperature vs cloud/hazes \citep{NeptunianTrends_CrossfieldKreidberg_2017,Kepler-51bd_FlatHSTspectrum_Libby-Roberts2020, HD3167c_HSTSpectrum_Guilluy_2021, TpCloudTrends_Estrela_2022} and planet mass vs metallicity \citep[[e.g.][]{Wakeford_MZ_2018AJ, MZTrend_Welbanks2019}.
Color-magnitude diagrams and trends regarding phase-curve properties and day-night temperature variations with equilibrium temperature have also been reported \citep[e.g.][]{Trends_BrownDwarfs&Exoplanets_Zhang_2020}.  Such population level studies are a start to understanding how atmospheric properties and planet composition relate to fundamental initial conditions, and are key to a full understanding of planet formation and evolution. However the lower temperature planets are poorly represented due to a paucity of atmospheric spectra in this regime.  This in turn is due to a combination of few strong candidates for spectroscopic follow-up and the intrinsic challenge in obtaining transmission spectra for cooler atmospheres (which will have smaller scale heights) at sufficient signal-to-noise ratio (SNR).

In this paper, we examine the capability of the upcoming \textit{Twinkle} space mission to advance the understanding of gaseous planets in the "cool" regime (which we define as being between 200 and 1000 K), through a dedicated spectroscopic survey consisting of a statistically meaningful sample of such planets.  The sample will cover a wide range of temperatures, and include planet sizes ranging from sub-Neptunes to Jovian planets. \textit{Twinkle} will obtain transmission spectra in the $0.5$-$4.5\mu$m wavelength range.  Such a survey would have the potential to verify and extend trends into the cool regime, providing key observational constraints for for atmospheric models and planet formation theories.

The paper is structured as follows: Section 2 outlines the scientific case for studying cool gaseous exoplanets, which is followed by a brief description of the \textit{Twinkle Space Telescope} in Section 3. Section 4 describes the construction of a preliminary candidate list based on known exoplanets spanning the parameter space of the cool gaseous planet population. Sections 5 and 6 describe two simulated studies performed on planets from the preliminary candidate list. Section 5 explores the ability of \textit{Twinkle} to constrain atmospheric metallicity and recover an injected metallicity trend, whilst Section 6 examines \textit{Twinkle}'s sensitivity to detecting clouds in gaseous planets over a range of sizes and temperatures.

\section{Cool gaseous planets}
\label{sec: Cool Gaseous Planets}

Spectroscopic observations of \enquote{cool} gaseous planets provide the opportunity to shed light on the physical and chemical processes that govern  H$_2$-He dominated atmospheres at low temperatures and the formation histories of this population.  

\subsection{Categorisation}
In this paper we sub-categorise the cool gaseous planets into nine sub-groups based on size and temperature.  A survey of cool gaseous planets across the full parameter space of size and temperature will allow the possibility of trends to be elucidated with respect to both parameters.  We limit the lower bound of temperature we consider to 200~K.  Below this level, the temperatures are generally beyond the outer limits of the "habitable zone", corresponding closer to those of the cold gas and ice giants of our solar system and in practice, there are hardly any transiting planets below this lower limit. 
Our upper temperature bound is 1000~K, which previous studies have used to define as boundary between hot Jupiters and cool giants \citep{Thorngren2019, Wallack2019}.  Avoiding terms like "temperate" or "warm" which have no consensus definitions,
we call all planets in this temperature range "cool".  We choose to further sub-divide the large cool temperature regime into three distinct temperature brackets: C1 (200-500~K), C2 (500-750~K) and C3 (750-1000~K) as distinctive patterns of chemical and physical processes are likely to occur with temperature. The C1 category would encompass planets in the "habitable zone". In terms of size, we include established planetary classes with radii above the Kepler radius valley mid-point ($1.75~R_\oplus$): sub-Neptunes ($1.75~R_\oplus-3~R_\oplus$), Neptunes ($3~R_\oplus-0.5~R_\text{J}$) and Jovians ($0.5~R_\text{J}-3~R_\text{J}$). These are planets where primary H$_2$-He dominated atmospheres are likely to be the norm \citep{RadiusValley_Fulton_2017, RadiusValley_CorePoweredLoss_Gupta_2019}. Their low mean molecular weights will mitigate some of the challenges associated with observing cool atmospheres, and hence they are favoured over rocky planets in this temperature regime for transmission spectroscopy.   

\subsection{The workings of cool atmospheres}
In the low temperature regime we would expect molecules such as NH$_3$ and CH$_4$ to dominate over N$_2$, CO$_2$ and CO in thermochemical equilibrium, with final abundances modulated by bulk elemental composition (higher metallicities tend to favour CO and particularly CO$_2$ over CH$_4$ and N$_2$ over NH$_3$) or the C/O ratio \citep{Madhusudhan2012, Moses2014}. However at lower temperatures reaction rates slow, such that the timescale for reaching chemical equilibrium increases compared to hotter atmospheres. Competing disequilibrium processes such as transport processes (e.g convection and eddy diffusion) and photochemistry would be expected to have stronger effects than in hotter atmospheres \citep{Prinn1977, Marley2014}. Molecules that tend to occur at low temperatures, like CH$_4$ and NH$_3$, are also more sensitive to photochemistry than their hot counterparts (CO and N$_2$) \citep{Moses2014}.
Such processes can change the composition, radiative balance and temperature-pressure profiles \citep{Moses2014}. HCN may be a significant molecule in cooler atmospheres as a result of coupled NH$_3$-CH$_4$ photochemistry, and CO may occur in the IR photosphere through CH$_4$-H$_2$O photochemistry or transport-induced quenching, but always at lower abundances than CH$_4$ \citep{Moses2014}.

The quench level is the atmospheric level where the chemical equilibrium time scale for a given reaction just falls below that for vertical (or horizontal) mixing, and the molecular abundances at this level are then transported to higher altitudes, resulting in a complex disequilibrium composition (also modulated by photochemistry) in the upper layers potentially probed in transmission.  In these upper layers, photochemical timescales may be expected to be shorter than chemical equilibrium so we might expect to see the byproducts of photochemistry at greater levels than for hotter atmospheres.
From breakdown of the photosensitive molecules CH$_4$ and NH$_3$, complex hydrocarbons and nitriles are more likely to occur in colder than in hotter atmospheres \citep{Moses2014}. This could include high-altitude photochemically-produced hazes which may modulate the energy balance of the planet. Photochemical reactions can also result in species that cause atmospheric warming and inversions or conversely could act as coolants (e.g. cooling by C$_2$H$_2$ and C$_2$H$_6$, byproducts of methane photolysis in the atmosphere of Jupiter).
Cloud condensates may form, reflecting the condensation profiles of low temperature molecules, giving rise to water, methane or ammonia clouds, which will also impact albedo and energy balance. The presence of clouds will also impact the measured quantities of the condensed species at higher altitudes reflected in the spectrum.  
The exact mechanisms and chemical pathways for disequilibrium chemistry remain an area of active research, as highlighted by the detection and subsequent interpretation of  SO$_2$ in the atmosphere of the hot Saturn WASP-39~b by the JWST Early Release Science (ERS) team \citep{2023_WASP39b_(SO2_photochemistry)_JWST/NIRSpec_Nature, 2023_WASP39b_(SO2)_JWST/NIRSpec-ERS_Nature1, 2023_WASP39b_(SO2)_JWST/NIRSpec-ERS_Nature_2}.
Therefore, while disequilibrium processes are expected and sophisticated atmospheric models exist to simulate the transport-induced quenching \citep[e.g][]{Moses2011, Drummond2020, Zamyatina2023} and photochemistry \citep[e.g.][]{Tsai2017} there are few if any observational constraints on the chemical kinetics from exoplanet observations. Further to this, there is a present lack of data on how planet and therefore gaseous envelope size would affect these processes.
A large and diverse spectroscopic survey of cool gaseous planets is ideally placed to find such "smoking guns".

 
\subsection{Planet formation quandaries}
Clues to the formation history of an exoplanet are encoded in its composition and therefore its atmospheric spectrum.  Elemental ratios, such as the C/O ratio, may be able to locate a planet origin location relative to different "ice lines" in the protoplanetary disk \citep[e.g.][]{CORatioinAtmospheres_Oberg_2011} and could be measurable in an exoplanet atmosphere.  The C/O ratio may be complicated by planet migration and/or planetesimal pollution. It may be possible to disentangle such pollution effects from origin location effects by examining a range of element ratios \citep[e.g.][]{Turrini2021, GiantPlanetFormation_ElementalRatios_Pacetti_2022}. Measurement of the atmospheric metallicity may also provide evidence for the formation mechanism. In core-accretion scenarios, the atmospheric metallicity is expected to be increased compared to the host star \citep{Thorngren_MetallicityDegeneracy_2016ApJ}, whereas in gravitational instability scenarios we would expect a near stellar metallicity. Current characterisations of the mass-metallicity relationship \citep[e.g][]{Wakeford_MZ_2018AJ, MZTrend_Welbanks2019} are supportive of core-accretion, but have large uncertainties and are derived mostly from hot giant exoplanets. Theoretical structural evolution models \citep{Thorngren2019} indicate that the mass-metallicity trend should continue in planets $<$~1000~K; however, there is currently limited data to test and confirm this.

Cool gaseous planets may present a challenge to planet formation theories. To hold onto a H$_2$-He atmospheres requires rapid formation of a massive core $\geq$ 10~$M_\oplus$ \citep[e.g.][]{CoreAccretion_Pollack_1996}. Traditional core accretion theory holds that such cores are more likely to form beyond the ice line where water ice adds to bulk and adhesion.  However many gaseous planets ranging from sub-Neptunes, through Neptune-sizes to Jupiter-sizes are found with equilibrium temperatures that would put them within the ice line. Indeed \cite{Hill2018} found $>$~70~planets of size $>$~3~$R_\oplus$ in the habitable zones of G, K and M-type stars, with occurrence rates ranging from 6 to 11.5\% depending on the stellar class. A planet formation quandary therefore exists in explaining the formation of these planets, requiring some modifications to basic core accretion models. Another problem is the presence of gas giants around M-dwarf stars (many of which are in the cool regime) \citep[e.g.][]{Kanodia2023}, where core accretion models predict slower accretion rates \citep[e.g.][]{Ida2005, Burn2021} that would make large core formation challenging on the timescale of the disc life time.  Gravitational stability is also a potential pathway to forming giant exoplanets \citep{GravitationalInstability_Boss_1997} and has seen renewed interest due to its ability to explain the existence of the growing number of M-dwarf gas giant exoplanets.
For gaseous planets within the water ice-line, formation scenarios include core-accretion beyond the water-ice line followed by disc-migration \citep{GiantPlanetFormationREVIEW_Paardekooper_2018}, or core formation interior to the water-ice line, followed by in-situ enrichment via gas, dust and pebble accretion close to the host star \citep{WarmGiantPlanetFormation_Knierim_2022}, along with other variations of the above models proposed in recent years.

Compositional information including atmospheric metallicity and elemental ratios such as the carbon-to-oxygen ratio (C/O) may therefore shed light on planetary formation and evolution processes in this regime.    

\subsection{Moons and habitability}
Giant planets have long been postulated to be likely exomoon hosts \citep{ExomoonFormation_Heller_2015, ExomoonPlanetaryMigration_Spalding_2016, ExomoonsTalk_Hill_2019, ExomoonsHIP41378f_Saillenfest_2023}, and although multiple exomoon candidates have been identified, there has yet to be a definitive detection \citep{ExomoonDetectability_Sucerquia_2020,ExomoonsThesis_Heller_2020,ExomoonOutgassingTransitDetection_Rovira_2021,ExomoonKeplerTTVs_Kipping_2023}. 
Though moons can form around planets of any size, cool gaseous planets may have a higher probability of hosting exomoons, owing to their having probably migrated inwards a shorter distance than hotter planets of a similar size and therefore being less likely to have undergone disruption the orbits of, or ejection of, their moons \citep{ExomoonPlanetaryMigration_Spalding_2016}. Furthermore, if such moons are sufficiently large, they themselves may have atmospheres, generated through mass transfer from their parent planet or, more likely, outgassing or volcanism as a result of tidal heating. Transmission spectroscopy may therefore provide a method by which exomoons could be detected around cool gaseous planets, with typical products of volcanism (including sodium- and potassium-bearing species) \citep{ExomoonOutgassingTransitDetection_Rovira_2021} unlikely to feature in cooler gaseous atmospheres. Thus cool giant spectroscopy may potentially yield evidence for the presence of exomoons. However, investigating this fascinating possibility requires a determination of the abundance of potential volcanic tracers and subsequently their detection feasibility with current and future instrumentation (eg:~\textit{Twinkle}, \textit{Ariel}, JWST and ELTs).  This is beyond the scope of the current paper and is left to future work.  We also note that such exomoons could be in the habitable zone in some cases,  and therefore could be locations for potential habitability. The habitability of cold giants themselves is an unexplored topic, and while liquid water layers or oceans such as those postulated for Hycean sub-Neptunes \citep{2021ApJ...MadhusudhanHycean} can be ruled out on the basis of pressure and temperature, the possibility of aerial biospheres could be explored. This has previously been raised in the context of Jupiter \citep{Sagan1976}, and more recently in brown dwarfs \citep{Yates2017,Lingam2019}, sub-Neptunes \citep{2021Univ....SeagerAerialBiospheres} and Venus \citep{Greaves2021}. 

\subsection{Previous observations}
Relatively few cool gaseous planets have been studied spectroscopically to date.
Sub-Neptunes are ubiquitous but small in size (reducing the $A_\text{p}$ factor in \autoref{eqn: Spectral Amplitude}), so despite the large number, there are few targets suitable for spectroscopic follow-up. Nonetheless, several sub-Neptunes in the "cool" regime have been studied spectroscopically. A non-exhaustive list includes: K2-18~b \citep{K2-18b_Tsiaras2019, K2-18b_Benneke2019, K2-18b_CH4_Bezard2020, K2-18b_CH4_DMS_Madhusudhan2023}, GJ~1214~b \citep{GJ1214b_Kreidberg2014, GJ1214b_JWSTspectrum_Gao_2023, GJ1214b_Kempton2023}, GJ~9827~d \citep{GJ9827d_HSTSpectrum_H2O_Roy_2023}, HD~3167~c \citep{HD3167c_HSTSpectrum_Guilluy_2021}, HD~97658~b \citep{HD97658b_HSTSpectrum_Knutson_2014} and TOI-270~d \citep{TOI-270d_HSTSpectrum_MikalEvans_2023}. Spectra of these planets have revealed greatly contrasting atmospheres. Observations conducted by HST, and more recently JWST, have shown the spectrum of GJ~1214~b to be flat well into the mid-infrared \citep{GJ1214b_Kreidberg2014, GJ1214b_JWSTspectrum_Gao_2023}, requiring significant cloud / haze production to explain, whilst HST spectra of K2-18~b showed clear absorption features at 1.4~$\mu$m relatively unimpeded by the presence of cloud. 

This feature has been inferred to be due to the presence of H$_2$O \citep{K2-18b_Tsiaras2019, K2-18b_Benneke2019} or CH$_4$ \citep{K2-18b_CH4_Bezard2020, K2-18b_CH4_Blain_2021}, with recent JWST/NIRISS and NIRSpec observations strongly detecting CH$_4$  \citep{K2-18b_CH4_DMS_Madhusudhan2023}. Absorption features at $\sim$1.4~$\mu$m, interpreted as due to H$_2$O have also been seen in HST spectra of GJ~9827~d, HD~3167~c and TOI-270~d, though observations spanning broader wavelengths are required to fully resolve the known degeneracy between H$_2$O and CH$_4$. 


Neptune to Jupiter-sized giant planets in the cool regime are also poorly characterised spectroscopically.  In terms of temperature, such planets provide a "missing link" between the two well-studied planetary populations of hot Jupiter exoplanets and the giants of our own Solar System.
Previous spectra of such planets include those for GJ~436~b \citep{GJ436b_Knutson2014, GJ436b_Hu&Seager2015}, GJ~3470~b \citep{GJ3470b_HSTSpectrum_Benneke_2019}, HD~106315~c \citep{HD3167c_HSTSpectrum_Guilluy_2021,HD106315c_Kreidberg_2022}, HIP~41378~f \citep{HIP41378f_TemperateJovian_Alam2022}, Kepler-51~b,~d \citep{Kepler-51bd_FlatHSTspectrum_Libby-Roberts2020}, K2-33~b \citep{K2-33b_HST+Spectrum_Thao_2023}, WASP-29~b \citep{WASP-29b|80b_HSTSpectrum_Wong_2022},  WASP-80~b \citep{WASP-80b_JWSTSpectrumNIRCam_Bell_2023}, and WASP-107~b \citep{WASP-107b_Kreidberg2018, WASP-107b_Piaulet2019_TalkAbstract, WASP-107b_Spake2021_AtmoEscape}, though many additional planets in this size regime have been the target of ground-based searches for metastable helium absorption \citep{MetastableHelium_Vissapragada_2022,MetastableHelium_SPIRou_Allart_2023}. 

Despite the increased number of spectra in the cool giant regime, these atmospheres remain poorly understood. Multiple planets, including HIP-41378~f, Kepler-51~b,~d, K2-33~b and WASP-29~b, exhibit spectra that are flat and featureless across the HST/WFC3 wavelength range, while others exhibit clear or muted absorption features at $\sim$1.4~$\mu$m.

\subsection{Surveys of cool gaseous planets}
There is a scarcity of systematic surveys of cool gaseous planets with high precision spectra. \cite{Kammer2015} obtained \textit{Spitzer} secondary eclipse measurements at at 3.6 and 4.5~\textmu m for five gas giants in the temperature range 980-1184~K.  They used the atmospheric CH$_4$/CO ratio as a marker of atmospheric metallicity, with results somewhat supportive of increased metallicity with lower masses. 
\cite{Wallack2019} performed a similar \textit{Spitzer} study on five further gas giants with $T_\text{eq} < 1000$~K. They found no evidence for a solar-system-like mass-metallicity relationship but did find a relationship between inferred CH$_4$/(CO+CO$_2$) and stellar metallicity.
More recently, \cite{Baxter2021} performed transmission photometry of 33 gaseous planets at 3.6 and 4.5~\textmu m using \textit{Spitzer}, of which 13 had temperatures between 500 and 1000~K. There was some evidence of a mass-metallicity relation: the cool planets ($<$ 1000~K) were generally biased with lower mass and appeared to have higher metallicity as well as lower eddy diffusion coefficients and a lack of methane compared to expectations. A lack of methane had previously been noted on a number of cool planets compared to equilibrium expectations constituting the so-called \enquote{missing methane problem}.  Methane has now recently been detected in two \enquote{cool} gaseous planets with JWST: K2-18~b and WASP-80~b \citep{K2-18b_CH4_DMS_Madhusudhan2023, Bell2023}.  More recently a Cycle 2 JWST survey of seven giant planets in the \enquote{cool} regime orbiting M-dwarf stars has been planned (JWST Proposal 3171, PI: S. Kanodia).

There is therefore a need for a homogeneous cool gaseous planet survey with wide wavelength coverage to further explore and constrain the relationship between planet mass and atmospheric metallicity, and open up this field of study. Cooler gaseous planets may provide more robust metallicity measurements than hotter gaseous planets as they will be significantly less affected by the degeneracy in radius between poorly-understood radius inflationary effects and increasing metallicity (and thus mean-molecular weight) which acts to suppress the atmospheric extent \citep{Thorngren_MetallicityDegeneracy_2016ApJ}. Improved metallicity measurements may also help to validate and extend reported mass-metallicity trends \citep{Wakeford_MZ_2018AJ,MZTrend_Welbanks2019} or support the absence of a trend \citep{2022arXiv_HSTTrends_70}. Furthermore, the temperature range spanned by cool gaseous planets is likely to aid in the exploration of cloud and haze coverage trends predicted from theoretical and laboratory work, hints of which have been seen in early HST observations \citep{Dymont_2022ApJ...937...90D, TpCloudTrends_Estrela_2022, SimilarSeven_McGruder_2023}.

Whilst the number of known cool giant planets is comparatively small, it has been steadily growing. In the last 18 months alone, \textit{TESS} photometry has lead to the discovery of intriguing cool giants such as the low-density warm-Jovian TOI-1420~b \citep{TOI-1420b_TESS_LowDensity_2023}, a warm Saturn TOI-199~b \citep{TOI-199b_TESS_CharacterisedwarmSaturn_2023} and several around M-dwarf stars \citep[e.g.][]{Mdwarf_TOI-3785b_Powers_2023,Mdwarf_TOI-904bc_Harris_2023, Mdwarf_TOI-5344b_Han_2023}. Re-observation of \textit{TESS} sectors during the second extended mission has enabled single transit and \enquote{duotransit} planetary candidates \citep{TESS_Duotransits_Hawthorn_2023}, many with orbital periods longer than $\sim$15~days, to be confirmed, with follow-up efforts by ground-based surveys such as the \textit{Next Generation Transit Survey} (NGTS) and the \textit{CHEOPS} space telescope subsequently hunting down and constraining the true orbital periods of these candidates. Examples of \enquote{cool} planets found in this manner include the sub-Neptunes HD~22946~d \citep{2023arXiv_HD22946bcd} and HD~15906~b and c \citep{2023arXiv_TESS+CHEOPS_HD15906}, the Neptunes HD~9618~b and c \citep{HIP9618_TESS+CHEOPS_2023}, TOI-5678~b \citep{TOI-5678b_TESS+CHEOPS+HARPS_2023} and the Jupiters TOI-4600~b and c \citep{TOI-4600bc_TESS+ground-based_2023} (planet c having a $T_\text{eq}$ of 191K).



However, compared to planets with hot atmospheres, cooler gaseous planets will have smaller scale heights and thus spectral features will give lower SNRs.  Co-adding of multiple transit observations is frequently used to improve SNR, but increases total observing time required. This is even more problematic for planets orbiting Sun-like stars where orbital distances are greater and periods longer. As such it is observationally more intensive to characterise cool gaseous planets (especially around Sun-like stars), and this is one reason why they are a challenging population. Dedicated spectroscopic surveys permitting repeat observations over several years would be key to opening up this population to detailed characterisation and obtaining sufficient homogenized samples for population-level studies. Two such dedicated surveys are planned in the coming decade: the ESA Ariel mission due for launch in 2029 and the \textit{Twinkle} mission which will precede it with a planned launch in 2025.   
 
\section{\textit{Twinkle}}
Developed commercially by Blue Skies Space Ltd. (BSSL), \textit{Twinkle} is a fast-tracked satellite based on heritage components, and that is expected to characterise many exoplanetary and solar system targets during its nominal 7-year lifetime \citep{Twinkle_Edwards_2019,Twinkle_SPIEPoster_2022}. Expected to launch in late 2025, the spacecraft will carry a 0.45-m diameter primary mirror, with an inner sanctum that is actively cooled to  <~90~K. \textit{Twinkle} has a spectrometer with two simultaneously operating channels across the visible and infrared: CH0 covering 0.5-2.4 $\mu$m at R $\leq$ 70 and CH1 covering 2.4-4.5 $\mu$m at R$\leq$50, each channel having its own grism element  \citep{Twinkle_SPIEPoster_2022}. \textit{Twinkle} will therefore expand on the total spectral wavelength coverage of HST WFC3 G102 (0.8-1.15 $\mu$m) and G141 (1.075-1.7 $\mu$m) gratings by a factor of $\sim 5$, whilst retaining similar spectral resolution. This will open up the opportunity to potentially break degeneracies between H$_2$O and CH$_4$ that are known exist in WFC3 observations, whilst also enabling the detection of strong absorption features from molecules such as CO$_2$ (2.0, 2.7, 4.3 $\mu$m), CO (2.34, 4.67 $\mu$m) and NH$_3$ (1.5, 2.0, 2.3, 3.0 $\mu$m) which lie outside the wavelength range covered by WFC3. 

\textit{Twinkle} has a field of regard centred on the anti-Sun vector encompassing $\pm 40^\circ$ about the ecliptic \citep{Twinkle_Edwards_2019}.  The primary exoplanet survey mission will take place in the first 3 years of operation. \textit{Twinkle} will therefore be uniquely positioned to provide homogeneous spectroscopic characterisation of a large number of exoplanetary atmospheres, something that will be challenging to achieve with JWST due to competition with other astrophysical disciplines for valuable telescope time. 

Launching several years prior to the European Space Agency's M4 mission, \textit{Ariel}, which has a 1-m class primary mirror and wavelength coverage from 0.5-7.8~$\mu$m \citep{ARIEL_Tinetti2022}, \textit{Twinkle} will additionally act as a useful precursor, observing many targets that fall within the current realisation of the Ariel target list over a substantially shared region of wavelength space. Consequently, insights gained from \textit{Twinkle} may be useful for informing future iterations of the \textit{Ariel} target list, allowing the combined science output of the two missions to be optimised. 

With the capability to provide the first large homogeneous survey of exoplanet atmospheres, we seek to explore \textit{Twinkle}'s ability to identify key molecules and elucidate trends in a population level study of cool gaseous planets, which we propose to integrate into the science plan of the \textit{Twinkle} Extrasolar Survey {\textcolor{blue}{(Twinkle collaboration, in prep.)}}. 

\section{Establishing a candidate list for the \textit{Twinkle} Cool Gaseous Planet Survey}
Candidate list construction for the proposed \textit{Twinkle} cool gaseous planet survey uses the database of confirmed planets from the NASA Exoplanet Archive\footnotemark[4]\footnotetext[4]{Planetary Systems Composite Table accessed 22/07/2022} to establish a preliminary target list of known planets. Transiting planets are selected based on three criteria: 
\begin{itemize}
    \item[1)] the existence of "transit" listed as the discovery method;
    \item[2)] the presence of a non-zero transit depth;
    \item[3)] the presence of a transit duration value,
\end{itemize}  
with any planet not meeting one or more of these criteria being filtered out. Planets with radii <~1.75~$R_\oplus$ are also removed, resulting in a list of transiting sub-Neptunian, Neptunian and Jovian class planets (see \autoref{tab: Transiting Planets by Characterisation}). To obtain the sample observable with \textit{Twinkle}, we perform a cut that eliminates planets with host stars outside $\pm 40^\circ$ of the ecliptic. The remaining 383 planets with radii between $1.75~R_\oplus$ and 3~$R_\text{J}$ and $T_\text{eq}$ between 200 to 1000~K form an initial \textit{Twinkle} cool gaseous planet candidate list.  The candidate list will be modified prior to launch as new discoveries are made. 

The initial candidate list is subjected to an SNR study, used to identify the number of transits required for each planet to achieve atmospheric detectability with \textit{Twinkle}. 
The findings of this study are then used to further filter the candidate list, leaving only planets that can be observed at or above the target SNR threshold during \textit{Twinkle}'s mission lifetime. This includes a cautious estimate of \textit{Twinkle}'s observing efficiency, scaling up the number of transits required to meet the SNR threshold by a uniform factor for each planet and resulting in our final lists of suitable and preferred candidates for the primary 3-year and extended 7-year \textit{Twinkle} exoplanet surveys.

\begin{table*}
    \caption{Population statistics for transiting exoplanets [derived from NASA Exoplanet Archive Planetary Systems Composite Table accessed 22/07/2022].
    Total numbers are shown in black, and the corresponding numbers of cool gaseous planets accessible within the field-of-regard (FOR) of the \textit{Twinkle} space telescope are shown bracketed in gray. The equilibrium temperatures for this table were obtained from the NASA archive where available or otherwise calculated from stellar and orbital parameters (assuming an albedo of 0.3).  At the time of submission, the number of cool gaseous planets in the \textit{Twinkle} FOR had increased from the 383 shown to 416.}
    \begin{tabular}{lr|ccc}
    \hline
    \multicolumn{1}{l|}{} & \multicolumn{1}{l|}{} & \multicolumn{1}{c|}{\begin{tabular}[c]{@{}c@{}}Sub-Neptunian\\ ($1.75~R_\oplus$ $\leq$ $R_\text{p}$ \textless~$3~R_\oplus$)\end{tabular}} & \multicolumn{1}{c|}{\begin{tabular}[c]{@{}c@{}}Neptunian\\ (3~$R_\oplus$ $\leq$ $R_\text{p}$ \textless~0.5~$R_\text{J}$)\end{tabular}} & \multicolumn{1}{c|}{\begin{tabular}[c]{@{}c@{}}Jovian\\ (0.5~$R_\text{J}$ $\leq$ $R_\text{p}$ \textless~3~$R_\text{J}$)\end{tabular}} \\ \hline \hline
    \multicolumn{1}{|l|}{Hot} & ($T_\text{eq} \geq$ 1000 K) & \multicolumn{1}{c|}{222}    & \multicolumn{1}{c|}{78}  & \multicolumn{1}{c|}{499}                                                                      \\ \hline
    \multicolumn{1}{|l|}{C3} &  (750 \textless\ $T_\text{eq}$ \textless 1000~K) & \multicolumn{1}{c|}{388 \textcolor{gray}{(75)}} & \multicolumn{1}{c|}{117 \textcolor{gray}{(30)}}  & \multicolumn{1}{c|}{84 \textcolor{gray}{(40)}} \\ \hline
    
    \multicolumn{1}{|l|}{C2} & (500 \textless\ $T_\text{eq}$ $\leq$ 750~K) & \multicolumn{1}{c|}{528 \textcolor{gray}{(109)}}    & \multicolumn{1}{c|}{166 \textcolor{gray}{(32)}}  & \multicolumn{1}{c|}{54 \textcolor{gray}{(22)}} \\ \hline
    
    \multicolumn{1}{|l|}{C1} & (200 \textless\ $T_\text{eq}$ $\leq$ 500 K) & \multicolumn{1}{c|}{309 \textcolor{gray}{(59)}}  & \multicolumn{1}{c|}{118 \textcolor{gray}{(7)}}  & \multicolumn{1}{c|}{75 \textcolor{gray}{(9)}}  \\ \hline

    \multicolumn{1}{|c|}{Cold} & ($T_\text{eq}$ $\leq$ 200 K) & \multicolumn{1}{c|}{1}  & \multicolumn{1}{c|}{3}  & \multicolumn{1}{c|}{6}  \\ \hline
    \end{tabular}
    \label{tab: Transiting Planets by Characterisation}
\end{table*}

\subsection{Establishing candidate planet SNR}
\label{sec: SNR Methodology determination}
Before examining the number of transits needed for each planets, we need to decide on a threshold SNR for spectral feature detection where the SNR is the ratio of the amplitude of a typical spectral feature $A_\text{p}$ to the noise on the transit depth (1$\sigma$ error bar) $\sigma_\text{p}(\lambda)$ at a given spectral binning.  If we assume a typical spectral feature corresponds to 5 scale heights, then we can use \autoref{eqn: Spectral Amplitude} to find $A_\text{p}$ with $n=5$, and $H$ for a given planet is obtained from \autoref{eqn: Scale Height}. The error bar for a given target SNR is given by
\begin{equation}
    \label{eqn: Fixed Errorbar Calculation}
    \centering
    |\text{error bar}| = \frac{2 \cdot 5 \cdot H \cdot R_\text{p}}{R_\text{s}^2} \cdot \frac{1}{\text{SNR}_\text{target}}
\end{equation}
We wished to verify that SNRs calculated this way corresponded to detectability of prominent molecules at high significance when simulated using an atmospheric radiative transfer code with parameters retrieved via Bayesian parameter estimation ("spectral retrieval"). The latter reflects the method of analysis that would be applied to a real observed transmission spectrum.  We decided to investigate nominal SNRs of 3 and 5.  
To this end a subset of the \textit{Twinkle} cool gaseous candidates (listed in \autoref{tab: Molecular Detectability Study Results}) have their atmospheres simulated using {\fontfamily{qcr}\selectfont TauREx3} (hereafter TauREx) \citep{TauRExI_2015ApJ, TauREx3CITATION_Al-Refaie_2021} as described further below.  For a given planet, in the calculation of $H$, $T_\text{eq}$ and $g$ are obtained and derived respectively from system values used by the \textit{Twinkle} radiometric tool, \textit{TwinkleRad}~\citep{Twinkle_SPIEPoster_2022}, based on \textit{ExoRad} \citep{ExoRad2_Mugnai_2022}, whilst $\mu$ is obtained directly from the TauREx atmospheric model. This allows calculation of $A_\text{p}$ for each planet, and error bars for SNRs of 3 and 5 obtained using \autoref{eqn: Scale Height}.

\begin{figure}
    \centering
    \includegraphics[scale=0.666]{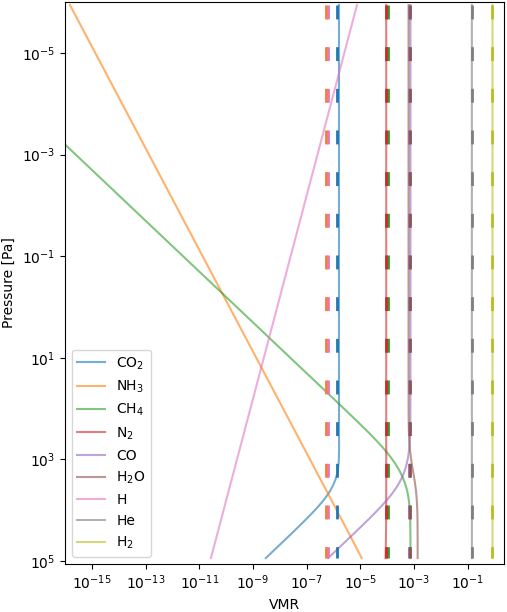}
    \caption{Modelled VMRs for HD~106315~c. Solid lines denote altitude-dependent chemical profiles under equilibrium conditions, whilst dashed vertical lines denote profile-averaged VMRs.}
    \label{fig:VMRs}
\end{figure}

\begin{figure*}
    \centering    
    \includegraphics[width=0.975\textwidth]{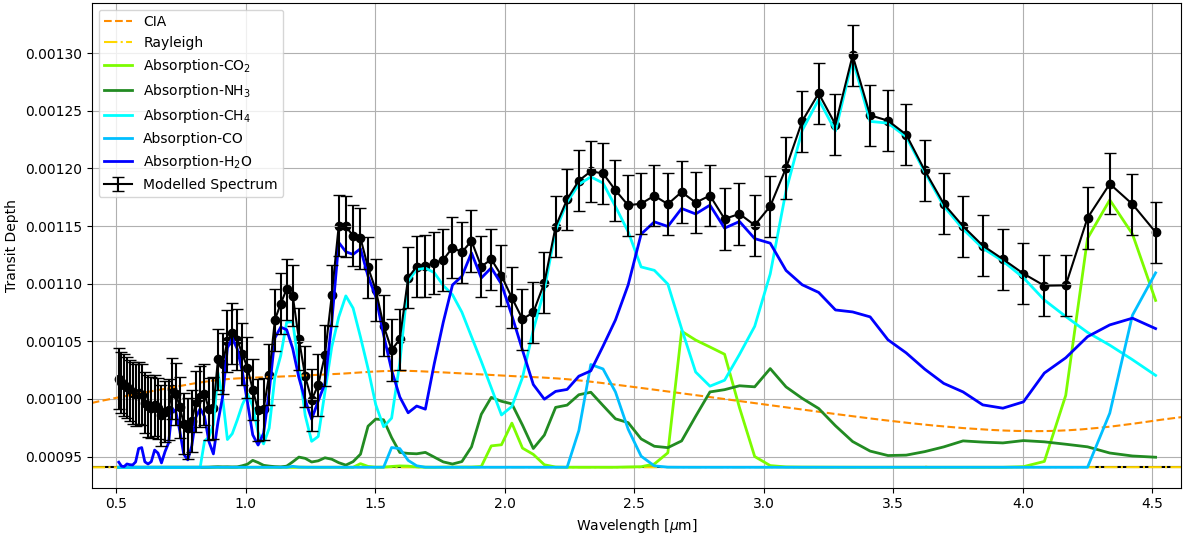}
    \captionsetup{justification=justified, width=0.975\textwidth}
    \captionof{figure}{Forward modelled spectrum for candidate HD~106315~c, binned to R=50 from 0.5-4.5 $\mu$m. Molecular absorption components are shown at the binned resolution of the final spectrum. Also shown are the total contributions to the final spectrum from CIA (between   H$_2$-H$_2$ and H$_2$-He) and Rayleigh scattering (all atmospheric molecules), with the final spectrum shown in black.} 
    \label{fig: HD106315c spectrum contributions}
\end{figure*}

For each planet, the atmospheric model and resulting model transmission spectrum are obtained as follows. TauREx was run initially using a model with 100 plane-parallel atmospheric layers spanning pressures from $10^{-6}$ to $10^5$~Pa, an isothermal temperature-pressure (T-P) profile at equilibrium temperature, $T_\text{eq}$, and equilibrium chemistry set by the {\fontfamily{qcr}\selectfont taurex\_ace} plugin based on the ACE equilibrium chemistry regime of \cite{ACEEq_2012A&A, ACEEq_2020A&A}, generated using solar C/O-ratio and metallicity values obtained for each planet using the trend found in \cite{MZTrend_Welbanks2019}. Altitude-dependent volume mixing ratios (VMRs) obtained this way for each molecule were then simplified to a single VMR value by taking the average across the profile (\autoref{fig:VMRs}) with this single value subsequently being used to set the free chemistry in the final model. The model was thus run again with the same initial conditions except this time with free chemistry, i.e. the fixed VMRs for the most abundant molecules (VMR $>$ $10^{-8}$), to give a final transmission spectrum.  We use opacity cross-sections from ExoMol \citep{ExoMol_Tennyson_2016} that include H$_2$O, CH$_4$, CO$_2$, CO and NH$_3$. 
In addition to molecular absorption, contributions from Rayleigh scattering and collision-induced absorption (CIA) between H$_2$-H$_2$ and H$_2$-He were included\footnotemark[5]\footnotetext[5]
{This work utilizes molecular cross-sections from 0.3-50 $\mu$m sampled at R=15000, which can be found at \url{https://www.dropbox.com/sh/13y33d02vh56jh2/AACh03L5h1QEbDYN7_-jMjBza/xsec/xsec_sampled_R15000_0.3-50?dl=0&subfolder_nav_tracking=1}. 
Rayleigh scattering data for all included atmospheric molecules and CIA files from HITRAN 2011 available at the above link are also used.}{}. 
To simulate an observed spectrum, the resulting near-local thermal equilibrium (near-LTE) cloud-free atmospheric spectra were then binned across the wavelength range covered by both \textit{Twinkle} spectroscopic channels to a fixed spectral resolution of $R$=50, approximating the performance of the instrument as detailed above.  To these binned points error bars were then added according to \autoref{eqn: Spectral Amplitude}, for the SNR~=~3 and SNR~=~5 cases.
An example of such a simulated observed spectrum is shown in \autoref{fig: HD106315c spectrum contributions} together with the different contributions to the spectrum.

\newpage
We note here that although disequilibrium chemistry processes and clouds and hazes are expected to be present in sub-$1000$ K planetary atmospheres \citep{WarmGiantPlanet_Chemistry_Fortney_2020,Dymont_2022ApJ...937...90D, WarmGiantExoplanets_Photochemistry_Fleury_2023}, such processes are poorly constrained at present, and hence the extent to which they may impact an individual planet cannot be evaluated with accuracy at this time.  We have therefore not attempted to include these processes in the atmospheric models created for this study, which should suffice for the purpose of establishing detectability of general spectral features.

Bayesian spectral retrievals are then conducted on the binned forward-modelled spectra produced using the nested sampling algorithm {\fontfamily{qcr}\selectfont nestle}\footnotemark[6]\footnotetext[6]{\url{https://github.com/kbarbary/nestle}} initiated in \enquote*{single} mode with 150 live points.  For each planet (with error bars corresponding to SNR~=~3 or SNR~=~5), the following retrievals are performed.
Firstly a baseline retrieval with the \enquote{full atmospheric} model, containing all atmospheric constituents with VMRs $>$ $10^{-8}$ (which generally included H$_2$O, CH$_4$, CO$_2$, CO and NH$_3$). 
The remaining three retrieval models are each initiated in the same manner as the full atmospheric model, but without one of either H$_2$O, CH$_4$ or CO. We select these molecules as they are found to consistently have the highest VMRs in our forward models. By comparing the Bayesian evidence obtained to that of the full atmospheric model, detection significances could be ascertained for each molecule. The detection significance of the three molecules is obtained via the log Bayes factor \citep{2008ConPh..49...71T, Benneke2013}, which is the difference in the log Bayesian evidence between full atmospheric model and the model with the molecule omitted. 

We find that for all the selected planets (\autoref{tab: Molecular Detectability Study Results}), CH$_4$ is detected at > 5$\sigma$ in all cases whether the error bars are derived from an SNR of 5 or 3.  With SNR=5, H$_2$O is detected to $\geq$ 5$\sigma$ in all cases, but at SNR= 3, water detection significance falls to a minimum of $\sim 3$ (\autoref{tab: Molecular Detectability Study Results}).  Our molecular detectability study also reveals that for the eight planets studied, CO is never detected above $2\sigma$, irrespective of SNR. Given the high VMR of this molecule in the forward models (log[VMR$_\text{CO}$] = -3 to -4), we attribute the weakness of detection to two factors. As can be seen in \autoref{fig: HD106315c spectrum contributions}, CO features at shorter wavelengths ($\sim$1.6 and 2.34~$\mu$m) are masked by spectral features from CH$_4$ and H$_2$O when all three molecules have similar atmospheric abundances, owing to CH$_4$ and H$_2$O having larger cross-sections. This challenge in robustly detecting CO is further compounded by the fact that the strongest observable band peaks at 4.7$\mu$m, just beyond the wavelength range covered by \textit{Twinkle} \citep{Twinkle_Edwards_2019}. 

Given that when we utliize error bars derived from an assumed SNR of 5 we obtain excellent detection significances for H$_2$O and CH$_4$ in all cases studied, we proceed by adopting SNR = 5 as the threshold to attain for all planets in the \textit{Twinkle} initial candidate list.

\subsection{Preliminary candidate list}
We next take the initial 383 candidates and estimate the number of transits needed in each case to reach an SNR of 5.  To calculate SNR we again find $A_\text{p}$ for each planet using \autoref{eqn: Spectral Amplitude}, but this time we obtain the transit depth errors from the radiometric tool, \textit{TwinkleRad}~\citep{Twinkle_SPIEPoster_2022} [via B. Edwards, private communication]. \textit{TwinkleRad} gives the 1$\sigma$ errorbar values on the transit depth for for a single transit. These values account for photon noise and instrumental effects and assume 100\% observing efficiency. The error bars from \textit{TwinkleRad} are given at the \enquote{native} wavelength grid of its \textit{Twinkle} model, which has a median resolving-power of 42 (ranging from 18-70). Model transmission spectra are thus binned to this native grid. 

\begin{table}
\caption{Detection significances (in $\sigma)$ of H$_2$O and CO obtained from retrievals conducted on simulated atmospheres for eight planets spanning the cool gaseous planets parameter space. Detection significance of CH$_4$ is  >5~$\sigma$ in all cases.}
\label{tab: Molecular Detectability Study Results}
\begin{tabular}{l|cccccc} 
\hline
\multicolumn{1}{c|}{\textbf{Planet name}} &
  \multicolumn{6}{c}{\textbf{Detection significance of individual molecules}}\\ \hline \hline
 & \multicolumn{2}{c}{Water (H$_2$O)}& \multicolumn{2}{c}{Carbon Monoxide (CO)}\\ 
 & SNR $\geq$ 3& SNR $\geq$ 5& SNR $\geq$ 3& SNR $\geq$ 5 \\ \hline
WASP-11 b &
  \multicolumn{1}{c}{4.4}  & \multicolumn{1}{c}{>5.0}  &
  \multicolumn{1}{c}{1.8}  & \multicolumn{1}{c}{1.7} \\ 
HD~63935 b &
  \multicolumn{1}{c}{>5.0}  & \multicolumn{1}{c}{>5.0}  &
  \multicolumn{1}{c}{1.7}  & \multicolumn{1}{c}{1.7} \\ 
HD~106315~c &
  \multicolumn{1}{c}{>5.0}  & \multicolumn{1}{c}{>5.0}  &
  \multicolumn{1}{c}{1.8}  & \multicolumn{1}{c}{1.3} \\ 
TOI-1130 c &
  \multicolumn{1}{c}{3.5}  & \multicolumn{1}{c}{>5.0}  &
  \multicolumn{1}{c}{1.8}  & \multicolumn{1}{c}{1.7} \\
GJ 436 b &
  \multicolumn{1}{c}{4.2}  & \multicolumn{1}{c}{>5.0}  &
  \multicolumn{1}{c}{1.7}  & \multicolumn{1}{c}{1.9} \\
HD~136352~c &
  \multicolumn{1}{c}{4.5}  & \multicolumn{1}{c}{>5.0}  &
  \multicolumn{1}{c}{1.6}  & \multicolumn{1}{c}{2.0} \\
AU~Mic~c &
  \multicolumn{1}{c}{3.1}  & \multicolumn{1}{c}{5.0}  &
  \multicolumn{1}{c}{1.8}  & \multicolumn{1}{c}{1.9} \\ 
TOI-178~g &
  \multicolumn{1}{c}{4.3}  & \multicolumn{1}{c}{>5.0}  &
  \multicolumn{1}{c}{2.0}  & \multicolumn{1}{c}{2.0} \\ \hline
\end{tabular}
\end{table}


Since the error on the transit depth is in reality wavelength dependent, so is the SNR.  However a single representative value was needed for a given planet for calculation of the number of transits. For this we use the lower quartile value for SNR across the full wavelength range covered by both channels.  
This ensures that $75\%$ or more of the spectrum achieves or exceeds the target SNR of the observation and that individual planets are not negatively biased by a single low-impact data point.

\begin{figure*}
    \centering
    \includegraphics[width=0.99\textwidth]{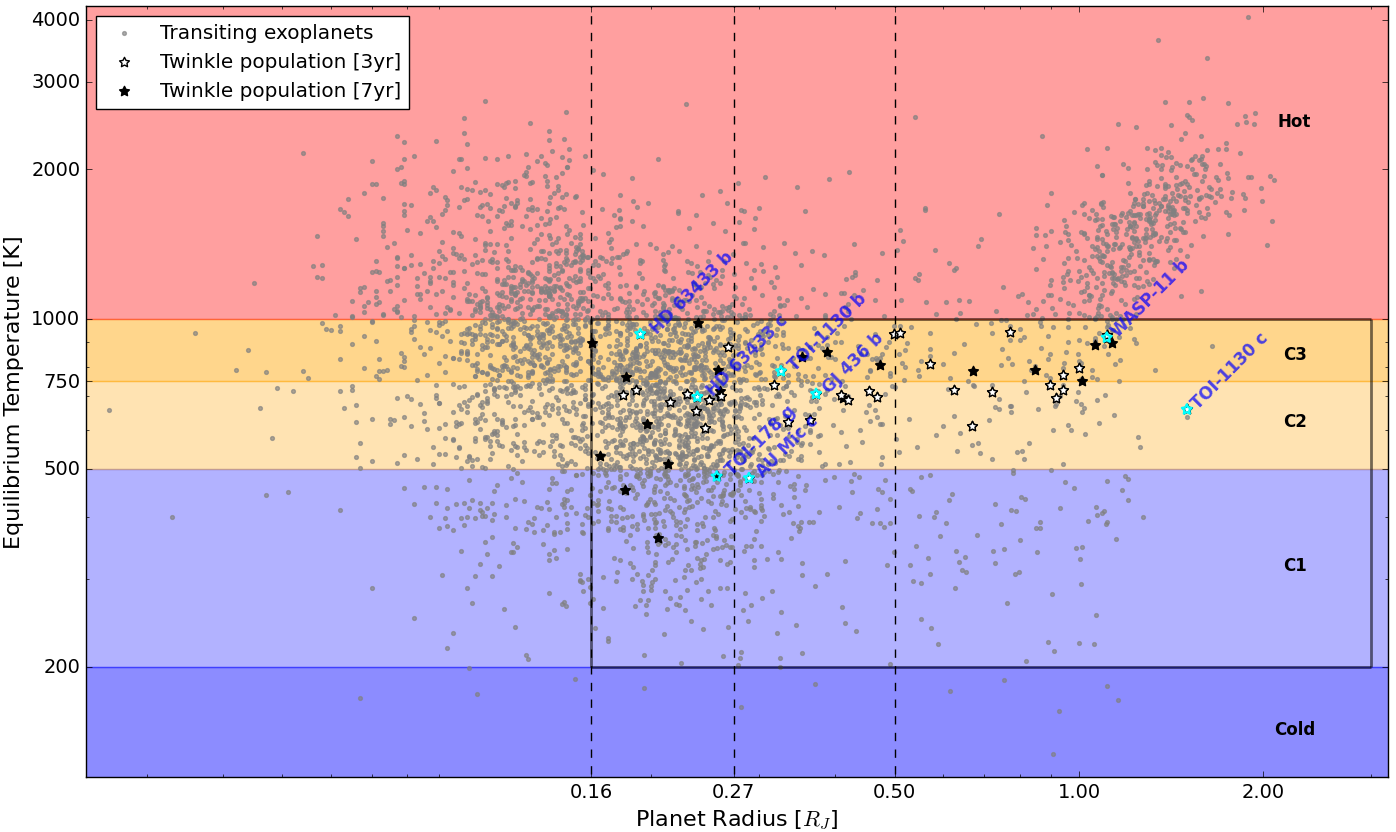}
    \captionof{figure}{Population of known exoplanets [NASA Exoplanet Archive, accessed 18/05/23], overlain by candidates targeted by this study for the primary 3-yr and extended 7-year \textit{Twinkle} exoplanet surveys. The different temperature regimes are indicated: Hot, C3, C2, C1 and cold.  C1, C2 and C3 are three sub-categories of the "cool" regime.  The bounded region indicated the parameter space of cool gaseous planets as defined in this paper.
    The equilibrium temperatures shown for \textit{Twinkle} 3- and 7-year mission candidates are obtained from methodologies outlined in \protect\cite{ARIELTargets_Edwards_2019} and used in the \textit{TwinkleRad} database. For all other planets, the equilibrium temperatures shown are obtained from the NASA archive where available or otherwise calculated from stellar and orbital parameters (assuming an albedo of 0.3).}
    \label{fig: Twinkle 7yr population}
\end{figure*}

In order to account for loss of data during Earth-occultation events that will arise due \textit{Twinkle}'s low-Earth orbit, we scale the \textit{TwinkleRad} error bars by a factor of 1/$\sqrt{0.75}$. This is done to simulate a conservative observing efficiency of 75\%, assumed to be the case for all planets within our observing sample.
Consequently, we re-calculate the representative single transit SNR of each planet, then obtain the number of transits, $N_\text{t}$, required to reach a threshold SNR of 5, rounded up to the nearest integer:  
\begin{equation}
    N_t = \left(\frac{5}{\text{SNR}_1}\right)^2
\end{equation}
where $\text{SNR}_1$ is the lower quartile SNR for a single transit. We combine this information with the orbital period for each planet to compute if the required number of transits could be observed within 3 or 7 years. This way we obtain a final candidate list of 36 and 57 planets for the primary (3-year) and extended (7-year) missions respectively. We show the distribution in the radius-temperature plane of the 57 candidates in the \textit{Twinkle} 7-year extended mission list in \autoref{fig: Twinkle 7yr population}. The final candidates are further separated into five distinct tiers using the following assigned criteria based on $N_\text{t}$ and the total integration time, $T_\text{int}$. The latter is calculated assuming a single transit observation lasts 3 $\times T_\text{14}$, where $T_\text{14}$ is the transit duration. The expectation is that a subset of higher tier candidates will be observed by \textit{Twinkle} during its lifetime.  

\begin{itemize}
    \item Tier 1: $N_\text{t}$ < 25 \textbf{and} $T_\text{int} \leq 25$ days
    \item Tier 2: $N_\text{t}$ < 50
    \item Tier 3: $N_\text{t}$ < 100
    \item Tier 4: $N_\text{t}$ < 150
    \item Tier 5: any other candidates
\end{itemize}

The candidates are listed with their $N_\text{t}$, $T_\text{int}$ and tierings in \autoref{tab: Twinkle 4yr and 7yr candidates_Tier_1_2_3} and \autoref{tab: Twinkle 4yr and 7yr candidates_Tier4_5}. These planets therefore give the preliminary candidate list for the \textit{Twinkle} cool gaseous survey and for the studies in this paper.  We note that 26/36 planets of the 3-year survey and 46/57 for the 7-year survey do not have any current transmission spectra.  However since this analysis was completed, the number of cool gaseous targets in the \textit{Twinkle} FOR has increased by 8\% (\autoref{tab: Transiting Planets by Characterisation}), and will continue to do so up to the time of the mission.  Our sample size thus represents a conservative value, with the final target list likely to be somewhat larger. While it is possible that some of these planets will be observed with JWST prior to the launch of \textit{Twinkle}, the \textit{Twinkle} survey being a homogeneous survey, inclusion of JWST planets in the sample would still be important.  We note that large numbers of transits are required for many planets in the sample which may ultimately not be practical in the real mission.  However, we expect Tier 1 planets (which range in $N_\text{t}$ from 1 to 22 transits) would be practical. Tier 2 planets may also be quite possible. Thus while the full sample shown here may not be ultimately adopted, a significant sub-sample of high-tier planets exists, covering a wide range of size and temperature, which would make a sizeable survey.


\section{Metallicity Trend detection study} 
\label{sec: Metallicity Trend}
One of the key properties for planet characterisation is metallicity, which is commonly split into two regimes: \textit{bulk}, describing the heavy-element content of the total planet and \textit{atmospheric}, pertaining to the atmosphere alone. As mentioned previously, atmospheric metallicity and elemental ratios can provide clues to the formation mechanism, location and migration history of the planet.
Atmospheric metallicity and elemental ratios, e.g. the C/O ratio, also set the initial elemental mix that controls the abundances of molecular species seen in thermochemical equilibrium.  The possible inverse relationship between planet mass and metallicity is consistent with core accretion scenarios. In this study we take the \textit{Twinkle} cool gaseous planet preliminary sample and investigate \textit{Twinkle}'s ability to elucidate a mass-metallicity trend in this sample, if one exists.

\subsection{Elucidation of a mass-metallicity trend}
We explore here whether an injected atmospheric metallicity trend can be recovered from a simulated atmospheric survey of cool gaseous planets using the \textit{Twinkle} 7-year candidate list (\autoref{tab: Twinkle 4yr and 7yr candidates_Tier_1_2_3} and \autoref{tab: Twinkle 4yr and 7yr candidates_Tier4_5}). The candidate list is well-suited for this investigation spanning  two orders of magnitude in mass. \\

We construct atmospheric forward models for each planet on the list, using {\fontfamily{qcr}\selectfont TauREx}, binning these to the native \textit{Twinkle} wavelength grid and adding re-scaled wavelength-dependent error bars obtained from \textit{TwinkleRad} to the resulting spectrum. The scaling accounts for the assumed observing efficiency of \textit{Twinkle} (75\%) and the unique number of transits, $N_\text{t}$, required for each planet in the sample to reach the desired SNR threshold of 5.  Our forward models have molecular abundances dictated solely by ACE equilibrium chemistry initialised with C/O=0.54 (solar) and a unique metallicity value for each planet obtained using the H$_2$O abundance mass-metallicity trend (including WASP-39 b) of \cite{Wakeford_MZ_2018AJ}. We use this reference since the metallicity values are given in units of solar metallicity, and the retrieved metallicities from TauREx are given in the same units.  Ten planets in the sample do not have currently measured masses but their estimated masses given in the NASA Exoplanet Archive are used for this purpose (these are planets for which there are no errors given in \autoref{tab: Twinkle 4yr and 7yr candidates_Tier_1_2_3} and \autoref{tab: Twinkle 4yr and 7yr candidates_Tier4_5}). 
 
Model atmospheres are again generated using a simple isothermal T-P profile spanning 100 plane-parallel layers $10^{-6}$ to $10^5$~Pa (10$^{-11}$~bar to 1 bar) and assume cloud-free conditions; however unlike the final models used in \autoref{sec: SNR Methodology determination}, we retain the full, altitude-dependent, VMR profiles for each chemical species. Transmission spectra are then generated from this atmospheric model by including molecular absorption
, CIA from both H$_2$-H$_2$ and H$_2$-He, and Rayleigh scattering. 
We subsequently perform Bayesian spectral self-retrievals to retrieve for atmospheric metallicity and C/O ratio. 
Each retrieval is initiated with the parameters and priors listed below in \autoref{tab: Metallicity Retrieval Setup}, with atmospheric metallicity and C/O retaining the same initialisation value and prior range for the full sample. 

Results are obtained for all candidates in the \textit{Twinkle} 7-year candidate list, with the exception of K2-138~f, which is subsequently.   
excluded from any analysis completed on the sample thereafter.  We found persistent errors halting the retrieval of K2-138~f.  While the exact cause of the error was not established, we note that K2-138~f has a very low mass, low gravity and an extremely high scale height of 1000~km, which could possibly be related to the computational failure. 

\begin{figure*}
    \centering
    \includegraphics[width=0.975\textwidth]{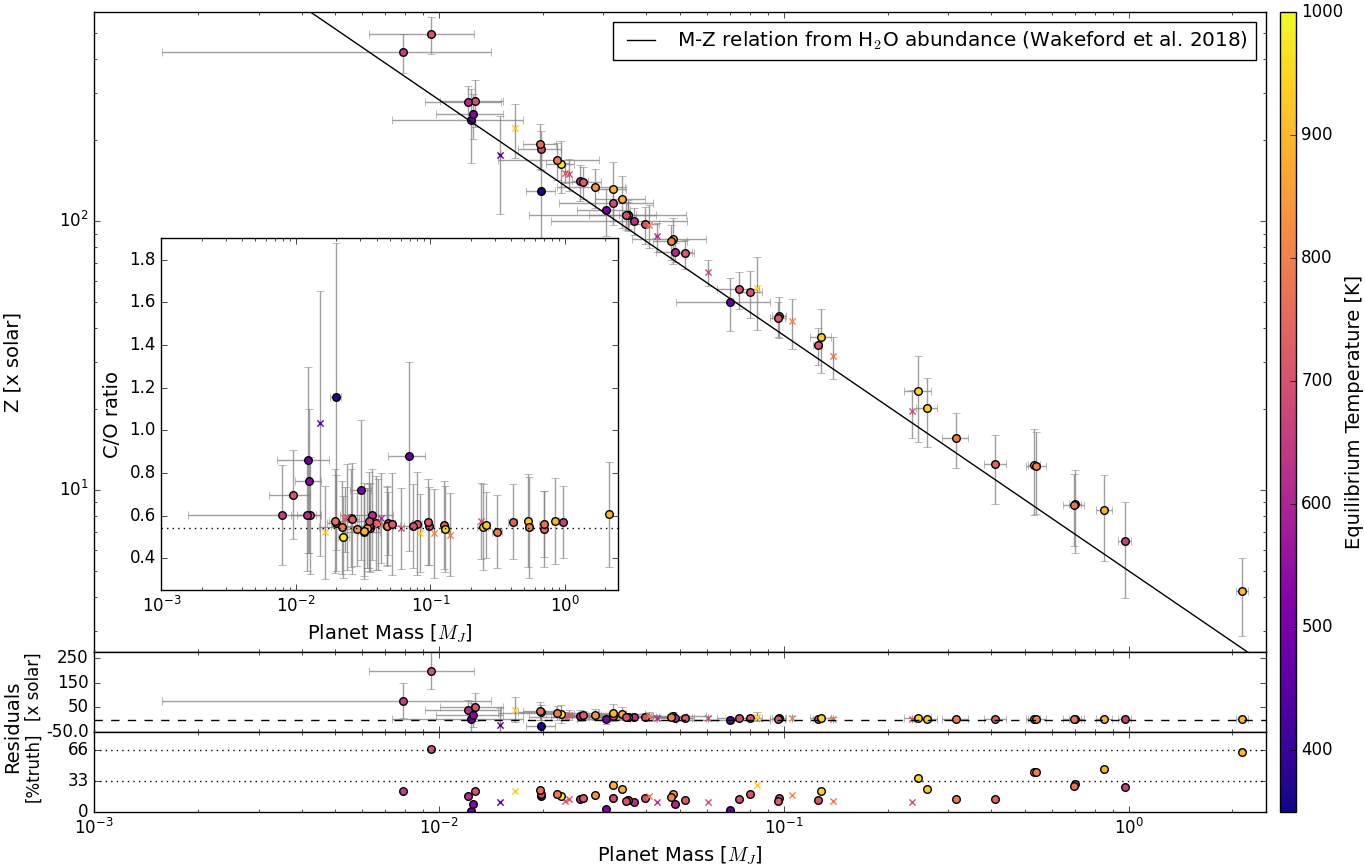}
    \caption{Retrieved metallicity plotted against literature mass. Inset shows retrieved C/O ratio against literature mass. Central points are the medians of the posterior distributions while error bars denoted the 1$\sigma$ confidence interval (16th-84th percentile ranges). Here we plot planets with currently measured masses as solid dots, whilst planets with masses estimated from M-R relations are plotted as crosses. Residuals with respect to input trend are shown in panel 2, whilst the residual normalised by the truth value is shown in panel 3.}
    \label{fig: Retrieved M-Z Trend}
\end{figure*}

\newpage
We plot the results obtained, along with their 1$\sigma$ error bars (mass from literature, metallicity from retrieval results) in \autoref{fig: Retrieved M-Z Trend}, along with the injected mass-metallicity trend.  In this figure the central points are the medians of the posterior distributions and the errors bars encompass the 16th-84th percentile ranges.
Although retrieved metallicity is found to be over-estimated in all but four cases across the sample, this is a small effect, with 51/56 planets (91\%) of the population results having the truth value within the $1\sigma$ confidence interval.  

This study indicates that \textit{Twinkle} has the capability of recovering a mass-metallicity trend in cloud-free 
cool gaseous planet atmospheres. However, we acknowledge that some planets may have hazes and clouds.  Also our assumption of equilibrium chemistry is likely a simplification given the unknown nature of such atmospheres and the likelihood of disequilibrium processes. Further studies could therefore examine the robustness of recovering a mass-metallicity trend under a more complex variety of atmospheric scenarios, including a mix of cloudy, cloudless and disequilibrium chemistry effects.
 
\begin{table}
    \caption{Initial values and prior ranges of free parameters in retrieval models used in the mass-metallicity study. Where factor bounds are used, values specified are multiplied by the truth value.}
    \centering
    \begin{tabular}{llll}
    \hline
    \textbf{Parameter} & \textbf{Input} & \textbf{Prior Type} & \textbf{Bounds}   \\ 
    \hline \hline
    Metallicity, Z     & 50             & Uniform, Linear   & [0.01, 750]         \\
    C/O ratio          & 0.54           & Uniform, Linear   & [0.1, 5.0]          \\
    Radius             & $R_\text{p}$   & Uniform, Factor   & [0.8~$R_\text{p}$, 1.2~$R_\text{p}$]        \\
    Temperature        & $T_\text{eq}$  & Uniform, Factor   & [0.8~$T_\text{eq}$, 1.2~$T_\text{eq}$]        \\ \hline
    \end{tabular}
    \label{tab: Metallicity Retrieval Setup}
\end{table}

\newpage
\subsection{C/O ratio retrieval}
In addition to retrieving for atmospheric metallicity, our study also retrieves for the carbon-to-oxygen ratio (C/O). Retrievals use the truth value as the input value in all cases, but implement a broad, uninformative prior as shown in \autoref{tab: Metallicity Retrieval Setup}. We find that the truth value is recovered within the 1$\sigma$ confidence interval in all cases (as shown in the inset in \autoref{fig: Retrieved M-Z Trend}). We do note that for 7 planets, the retrieved median values deviate strongly (by $\geq$ 0.135) 
from the injected truth, with large error bars. We find that in the majority of these deviated cases, the posterior distributions were asymmetric with an extended positive tail, which may in part explain the over-estimation given by the median.

\newpage
\subsection{Exploring sources of bias in metallicity retrievals}
\begin{figure}
    \centering
    \includegraphics[width=0.92\columnwidth]{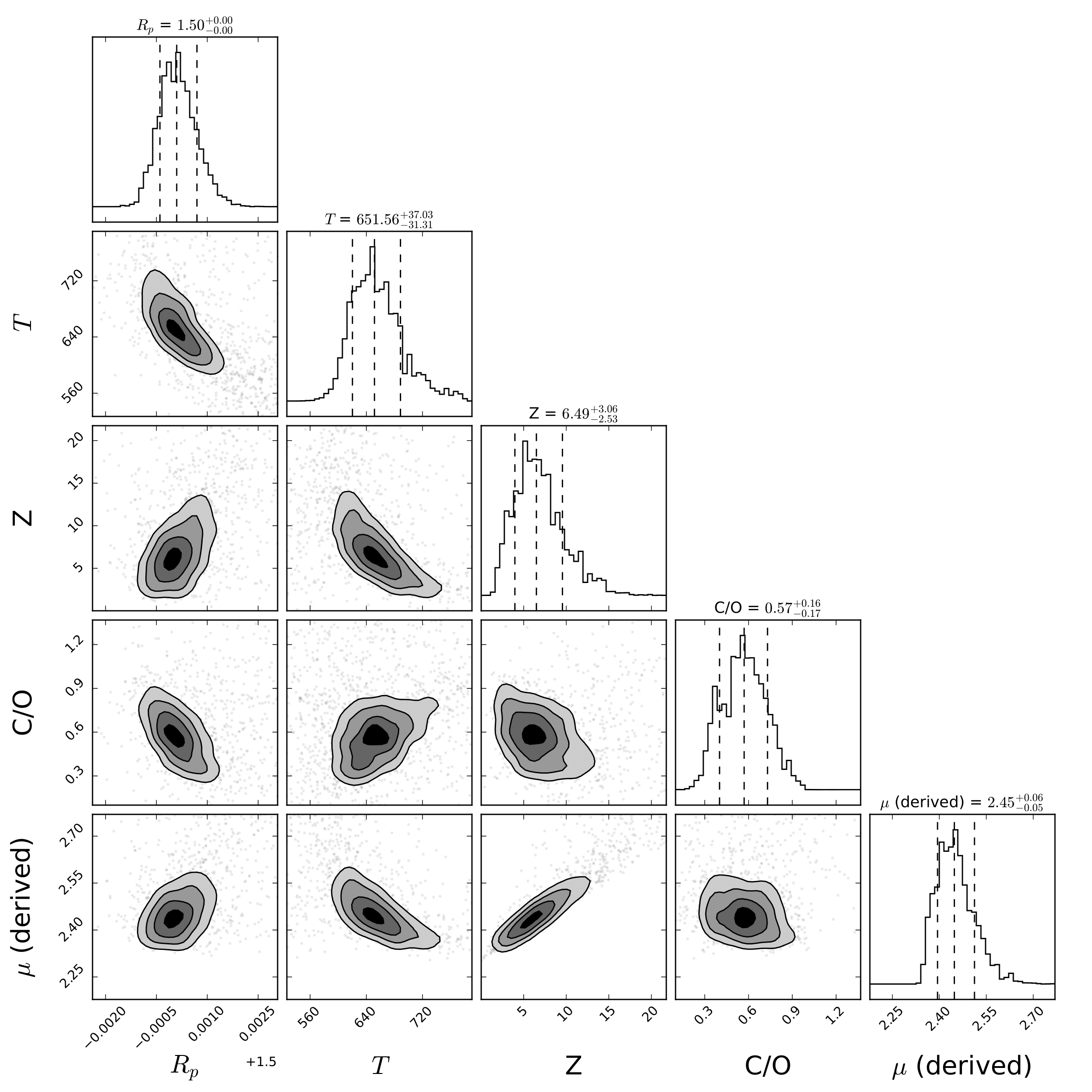}
    \captionof{figure}{Retrieved joint posterior distribution for Tier 1 candidate TOI-1130~c.}
    \label{fig: TOI-1130c metallicity retrieval}
\end{figure}

To explore the slight over-estimation in retrieved metallicity seen for the bulk of the \textit{Twinkle} 7yr candidate population, we first explore the posterior distributions generated from the nested sampling retrieval results. Although there is some evidence of non-Gaussianity in the posterior distributions, as shown for TOI-1130~c in \autoref{fig: TOI-1130c metallicity retrieval}, we find no evidence for systematic over-estimation of the median due to effects of sampling an asymmetric distribution. 

We therefore elect to investigate the effect, if any, that the modelling and retrieval process has in creating this bias, by varying combinations of the wavelength grid and spectral resolution of the modelled spectra. This approach is taken as atmospheres are initially generated in TauREx using cross sections that span the wavelength range 0.3-50~$\mu$m at a spectral resolution of $R=15000$, resulting in a substantial loss of information from the model inputs when spectra are binned down to the specifications of the observing instrument.  Three additional sets of models (\autoref{tab: Metallicity Trend Study Cases}, cases 2, 3 and 4) are generated using the wavelength ranges and spectral resolutions as listed, for each planet in the \textit{Twinkle} 7-year candidate list. 

\begin{table*}
\caption{Metallicity bias study test cases.}
\label{tab: Metallicity Trend Study Cases}
\begin{tabular}{lll}
\hline
\textbf{Case Number \& Name} &
  \textbf{Wavelength Range} &
  \textbf{Wavelength Grid} \\ \hline \hline
1 \textit{Twinkle} native & 0.5-4.5~$\mu$m & \textit{Twinkle} native (average $R$ $\sim$ 42) \\
2 \textit{Twinkle} - HST WFC3/G141 range & 1.1-1.7~$\mu$m & \textit{Twinkle} native (average $R$ $\sim$ 44) \\
3 \textit{Twinkle} $R=70$ & 0.5-4.5~$\mu$m & fixed $R=70$     \\
4 \textit{Ariel} $R=70$   & 0.5-7.8~$\mu$m & fixed $R=70$     \\ \hline
\end{tabular}
\end{table*}

\begin{figure*}
    \centering
    \includegraphics[scale=0.63]{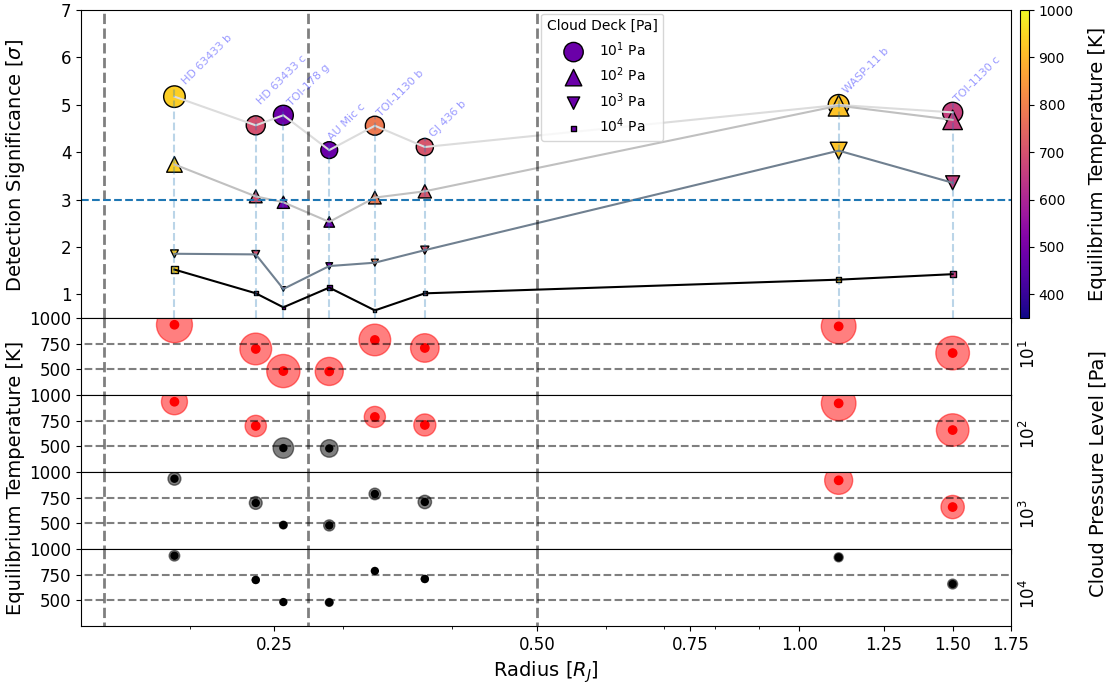}
    \caption{The upper plot shows statistical significance of cloud detection, derived from the Bayes factor preference for a model with clouds over model without clouds, for eight planets. The lower plot shows the temperature-radius plane for those eight planets with red circles indicating cases where detection significance is $>3\sigma$ while black circles denote cases at  $<3\sigma$.}
    \label{fig: Cloud Study Results}
\end{figure*}

Here, case 1 is considered to be our baseline case, representing the performance of \textit{Twinkle} based on current knowledge and modelling. We base our case 2 model on the widely-used HST WFC3/G141 instrument configuration to examine the effect of reducing the wavelength coverage compared to case 1 on retrieved metallicity, retaining the wavelength resolution grid of \textit{Twinkle} such that any changes in retrieved metallicity can be attributed solely to the reduction in wavelength coverage.  We use cases 3 and 4 to see the effect of increasing wavelength coverage from that of \textit{Twinkle} to that of Ariel (approximately doubling the wavelength range).  For consistency cases~3~and~4 utilize a fixed $R$ of 70.  This $R$ value does not reflect the true $R$ for \textit{Twinkle} or Ariel (which will vary will wavelength), but is chosen as a nominal value for the purposes of this comparison.
Increased wavelength coverage would be expected to boost the sensitivity to of the retrieval to molecules such as CO (see \autoref{sec: SNR Methodology determination}). This approach is taken rather than accurately modelling an observation of these targets with \textit{Ariel} due to the differences in native spectral resolution between \textit{Twinkle} and \textit{Ariel}, which would inhibit the ability to isolate the dependence on wavelength coverage on any systematic findings. We keep the error bars on the spectra the same for each individual planet across all cases, and perform self-retrievals on the semi-physically-motivated models of cases 2, 3 and 4 in the same manner described in \autoref{sec: Metallicity Trend}. 

Our findings are as follows: for case 2, average retrieved 1$\sigma$ error ranges are just over a factor of 3 higher for atmospheric metallicity compared to case 1, yet despite this, values are typically found to be proportionately more over-estimated, with only 35/56 planets (62.5\% of the population) having the truth value within the $1\sigma$ confidence interval.
Comparisons between the baseline case (case 1) and case 3 show that the average retrieval error bars are comparable; the average 1$\sigma$ confidence interval is 5\% greater in case 3 compared to case 1.  The number of planets with truth values within the 1$\sigma$ range is the same. Similar results are seen when comparing cases 3 and 4: in both cases 51/56 planets have truth values within 1$\sigma$ of the retrieved atmospheric metallicity values. However, the average retrieved 1$\sigma$ range is 10\% smaller in case 4 compared to case 3. 

Importantly, we find that in a little over two thirds of the sample (38/56 planets), case 4 spectra, with broader wavelength coverage, yield metallicity values that lie closer to the truth than in case 3. While this suggests that increased wavelength coverage leads to an improvement in the ability to recover atmospheric metallicity from spectra, further investigation shows that typically the absolute difference between retrieved metallicity and the truth value,$|Z_\text{ret} - Z_\text{truth}|$, is only marginally different between cases 3 and 4. We calculate this difference for each of the 56 planets in our sample using: 
\begin{equation}
\label{eqn: Metallicity Case Comparison}
    \delta = |Z_\text{ret} - Z_\text{truth}|_3 - |Z_\text{ret} - Z_\text{truth}|_4
\end{equation}
 
The average value for $\delta$ is 1.3, which is well within the quadrature sum of the 1$\sigma$ errors of the retrieved metallicities for each case (where the 1$\sigma$ error is taken to be half of the 1$\sigma$  confidence interval which is itself the 16th-84th percentile range).
This suggests that the differences between cases 4 and 3 are not significant.
We conclude that the slight bias seen in retrieved metallicity is likely at least in part due to information loss from reduced wavelength coverage, although other possible causes may exist.

\section{Cloud Detection Study}

Clouds are thought to be present to some extent in the atmospheres of all planets \citep{Clouds_Helling_2021} and are known to make a significant contribution to the overall planetary albedo, a key parameter in the determination of equilibrium temperature \citep{TpCloudTrends_Estrela_2022}. Capable of influencing temperature, and hence the energy budget of an atmosphere, clouds are intrinsically coupled to T-P structure of the atmosphere and its chemical composition   
\citep{ExoplanetAtmospheresREVIEW_Madhusudhan_2016}, with clouds of varying compositions expected to form as atmospheric species cross condensation fronts.  Although the basic mechanisms by which cloud formation occurs are well-formulated, the precise details and balance of pathways are currently debated \citep{Clouds_Helling_2019, Clouds_Helling_2021}.
Consequently, the detection of clouds in in cool gaseous planets and subsequent inference of the altitudes at which they are present, has the potential to provide insight into atmospheric processes governing cloud formation in this regime.

We therefore investigate \textit{Twinkle}'s capability to detect global cloud layers for a sub-sample of planets taken from the candidate list (\autoref{tab: Twinkle 4yr and 7yr candidates_Tier_1_2_3} and \autoref{tab: Twinkle 4yr and 7yr candidates_Tier4_5}).  We selected one representative planet from each of the nine major sub-divisions shown in \autoref{fig: Twinkle 7yr population} with the exception of C1-Jovian planets for which no examples exist in our list. The selected planets are indicated by name in \autoref{fig: Twinkle 7yr population}.
 
Our goal here is to find constraints on the pressure levels at which cloudy atmospheres can be distinguished from cloud-free ones, functioning also as a top-level search for potential trends in cloud detectability across the different planetary types. 
Forward models are constructed using TauREx for each of the eight planets selected. Models are generated as previously described with isothermal profiles. ACE equilibrium chemistry is initiated with C/O=0.54 (solar) and unique metallicity values for each planet from the H$_2$O mass-metallicity trend (including WASP-39 b) of \cite{Wakeford_MZ_2018AJ}.  The resulting altitude-dependent chemical profiles are then simplified to an individual altitude-independent VMR as described in \autoref{sec: SNR Methodology determination} for H$_2$O, CH$_4$, CO$_2$, CO, NH$_3$ and N$_2$, and the model run again with fixed VMRs to generate the final transmission spectra.  We include opacity contributions from molecules, Rayleigh scattering and CIA.
In addition, an optically thick grey cloud layer is modelled at a given pressure level, which is varied between one of four pressures: 10$^{4}$, 10$^3$, 10$^2$ and 10 Pa in different runs.
In each case, the resulting forward model spectra are then subjected to two Bayesian spectral retrievals. Both retrievals are conducted with identical initial input parameters and bounds, retrieving for all molecule VMRs, temperature and planet radius, with one retrieval model retrieving for cloud pressure level, while the other does not include clouds.

Planetary radius and equilibrium temperature are fit with the same priors from \autoref{tab: Metallicity Retrieval Setup}, whilst log-uniform priors with bounds [$10^{-12}$,~$10^{-1}$] are used for all molecule VMRs (H$_2$O, CH$_4$, CO$_2$, CO, NH$_3$ and N$_2$). Where clouds are included in the retrieval model, log-uniform priors with bounds [$10^{-6}$,~$10^{5}$]~Pa are used. We use the difference in the log Bayesian evidence from each pair of retrievals to determine a detection significance \citep{2008ConPh..49...71T, Benneke2013}  for cloud (at the given pressure level) in each case. The results are illustrated in \autoref{fig: Cloud Study Results}. 

We find that high-altitude clouds at 10~Pa are detectable above $3\sigma$ irrespective of the planetary or temperature regime they are in. Deeper clouds at 100~Pa can be detected above $3\sigma$ across all planet sizes, but only in C3 (500-1000~K) planetary atmospheres. 

At cloud pressure of 1000~Pa, we obtain $3\sigma$ detections only in the two Jovian planets, suggesting it may be possible to probe physical processes and T-P structure in the deeper layers down to 0.01 bar.  At $10^4$ Pa (0.1 bar), we are unable to distinguish cloudy and cloud-free atmospheres at $\geq$ 3$\sigma$ in any of the cases.  

These results indicate that \textit{Twinkle} will be unlikely to detect clouds deeper than 0.1 bar in any cool gaseous planets, and should be able to detect clouds at 0.0001 bar (10 Pa) or higher in all cases.  We find a rough indication of a trend of improved detectability with temperature and size of the planet. We also observe a tentative trend of decreasing sensitivity with planetary size across the sub-Neptune group and sub-Neptune-Neptune boundary; however this may also be explicable by reduction in temperature in the lower detection planets.  The robustness of these conclusions is limited due to the small sub-sample size used here. These initial results can be built on by further investigation using a larger sub-sample or a large grid of completely simulated planets (providing exact temperature and radius controls).

\section{Conclusions}
In summary, we present a first realisation of a tiered candidate list for the proposed \textit{Twinkle} cool gaseous planets survey, based on currently confirmed exoplanets. We find that \textit{Twinkle} has the potential to characterise the atmospheres of up to 36 and 57 planets in the primary 3-year and extended 7-year survey respectively. Candidates identified include 27 and 46 planets respectively that, to the best of the authors' knowledge at the time of submission, do not have precise transmission spectra.  A survey using just Tier~1 and Tier~2 planets, would yield 20 planets ranging in mass from 0.0079 to 0.97~$M_\text{J}$ and 480 to 941~K in temperature.  Due to growing numbers of discoveries in the cool gaseous regime, e.g. by \textit{TESS}, these sample sizes can be considered conservative and will be greater by the time of \textit{Twinkle}'s launch. The final candidate list used for the \textit{Twinkle} mission will be updated to include new discoveries

\textit{Twinkle} is well-positioned to provide the first opportunity to garner insights into cool gaseous planets at a population level.  The 3-year baseline survey includes all 15 Tier~1 candidates (7~Jovians, 4~Neptunians, 4~sub-Neptunians, collectively spanning $\sim$480-934~K) and all 5 Tier~2 candidates.  These planets will be the highest priority candidates for the survey. If planets up to Tier~3 are included, a 3-year survey would have 31~planets, and a 7-year survey would have 35~planets.  


Our study predicts that major molecular species expected to be present under cloud-free near-equilibrium conditions in sub-1000~K atmospheres can be detected to high significance, and that if present, trends in atmospheric metallicity can be reliably identified across the sample of surveyed planets. The C/O ratio was also recovered to within 1$\sigma$. 
These studies show that \textit{Twinkle} will provide valuable data capable of contributing towards the understanding of trends within the cool gaseous planet demographic and beyond, as shown in \autoref{fig: Retrieved M-Z Trend}. Such data has the potential to inform atmospheric models and further planet formation theories for giant planets in this understudied temperature regime.

We also predict that based on a simple grey cloud model, \textit{Twinkle} has the potential to detect high-altitude cloud decks at or above 10~Pa in all atmospheric spectra in the cool regime, but would be insensitive to clouds below $10^4$~Pa in all cases.  We find a rough and tentative trend in cloud detectability with planetary temperature and size.

The cool gaseous planets represent a new frontier in exoplanet science, promising to expand our understanding of atmospheric physics and chemistry, as well as planet formation and evolution.  The \textit{Twinkle} cool gaseous planet survey has the potential to open up this uncharted territory of exoplanet parameter space with paradigm-shifting results.
20 additional survey candidates present across tiers 2 and 3, and many new candidates, typically discovered around bright stars, hence amenable for transmission spectroscopy, confirmed and being actively refined by \textit{TESS} and other facilities (including 32 new cool gaseous planets in the Twinkle field-of-regard) this proposed survey to be conducted by \textit{Twinkle} promises to deliver a small, but statistically meaningful sample of homogeneous spectra.


\section*{Acknowledgements}
We thank Ben Wilcock of Blue Skies Space Ltd. for his comments on the manuscript.  L.B. is supported by a STFC doctoral training grant. Additionally, we thank the anonymous reviewer for their helpful comments.

\section*{Data Availability}
The data underlying this article will be shared on reasonable request to the corresponding author.

\bibliographystyle{mnras}
\bibliography{references} 



\appendix
\section{Candidate Lists}
\begin{table*}
    \caption{Candidate planets in Tiers $1$, $2$ and $3$ for the \textit{Twinkle} cool gaseous planet survey. Planetary and stellar parameters listed are used throughout this work and were obtained from the NASA Exoplanet Archive or calculated based on assumptions presented in \protect\cite{ARIELTargets_Edwards_2019}. Where unavailable in the archive, host star spectral type uses SIMBAD values ($^\ddag$) or estimation based on comparable stars ($^\dagger$).}
    \begin{tabular}{@{}l|rrrlc|cccc|ccc@{}}
    \toprule
      \textbf{Planet name} &
      \textbf{$\bm{N_\text{t}}$} &
      \textbf{$\bm{T_\text{int}}$} &
      \textbf{$\bm{R_\text{p}}$} & \textbf{$\bm{M_\text{p}}$} &
      \textbf{$\bm{T_\text{eq}}$} &
      \textbf{In 3yr} &
      \textbf{In 7yr} &
      \textbf{Tier} & 
      \textbf{Transmission} & \textbf{$\bm{R_\text{s}}$} & \textbf{$\bm{T_\text{s}}$} & \textbf{Spectral}\\ 
      & & \textbf{[days]} & \textbf{[\bm{$R_\text{J}$}]} & \textbf{[\bm{$M_\text{J}$}]} & \textbf{[K]} & \textbf{List} & \textbf{List} & & \textbf{Spectra} & \textbf{[$\bm{R_\odot}$]}& \textbf{[K]} & \textbf{Type}\\ \hline \hline
    AU Mic b    & 2    & 0.904   & 0.363 & $0.0368_{-0.0157}^{+0.0157}$ & 626.428  & Y & Y & 1 & False & 0.75 & 3700 & M1 \\
    AU Mic c    & 20   & 11.359  & 0.289 & $0.0699_{-0.0211}^{+0.0211}$ & 479.593  & Y & Y & 1 & False & 0.75 & 3700 & M1 \\
    GJ 1214 b   & 15   & 1.581   & 0.245 & $0.0257_{-0.0014}^{+0.0014}$ & 603.949  & Y & Y & 1 & True & 0.21 & 3250 & M4 \\
    GJ 3470 b   & 7    & 1.602   & 0.408 & $0.0396_{-0.0040}^{+0.0040}$ & 702.732  & Y & Y & 1 & True & 0.55 & 3600 & M1.5 \\
    GJ 436 b    & 4    & 0.590   & 0.372 & $0.0799_{-0.0063}^{+0.0066}$ & 708.023  & Y & Y & 1 & True & 0.46 & 3586 & M2.5\\
    HD 136352 c & 17   & 6.975   & 0.260 & $0.0354_{-0.0020}^{+0.0021}$ & 701.029  & Y & Y & 1 & False & 1.06 & 5564 & G4 \\
    HD 63433 b  & 20   & 8.014   & 0.192 & $0.0166$ & 934.262  & Y & Y & 1 & False & 0.91 & 5640 & G5 \\
    HD 63433 c  & 17   & 8.584   & 0.238 & $0.0239$ & 698.391  & Y & Y & 1 & False & 0.91 & 5640 & G5 \\
    K2-141 c    & 1    & 0.333   & 0.624 & $0.0233$ & 720.185  & Y & Y & 1 & False & 0.68 & 4599 & K7 \\
    K2-24 c     & 22   & 19.008  & 0.669 & $0.0485_{-0.0057}^{+0.0060}$ & 610.027  & Y & Y & 1 & False & 1.16 & 5625 & G9 \\
    TOI-1130 c  & 8    & 2.056   & 1.500 & $0.9740_{-0.0440}^{+0.0430}$ & 658.620  & Y & Y & 1 & False & 0.69 & 4250 & K7 \\
    V1298 Tau b & 22   & 18.094  & 0.916 & $0.2360$ & 695.412  & Y & Y & 1 & False & 1.34 & 4970 & K0 \\
    WASP-107 b  & 1    & 0.351   & 0.940 & $0.0960_{-0.0050}^{+0.0050}$ & 719.812  & Y & Y & 1 & True & 0.67 & 4425 & K6 \\
    WASP-69 b   & 2    & 0.556   & 1.110 & $0.2600_{-0.0185}^{+0.0185}$ & 928.605  & Y & Y & 1 & True & 0.86 & 4700 & K5 \\
    WASP-80 b   & 10   & 2.695   & 0.999 & $0.5400_{-0.0350}^{+0.0360}$ & 799.369  & Y & Y & 1 & True & 0.59 & 4143 & K7-M0 \\ \hline

    HD 183579 b & 42   & 23.804  & 0.317 & $0.0352_{-0.0170}^{+0.0170}$ & 736.129  & Y & Y & 2 & False & 0.99 & 5788 & G2 \\
    TOI-1064 c  & 36   & 10.697  & 0.237 & $0.0079_{-0.0057}^{+0.0063}$ & 653.762  & Y & Y & 2 & False & 0.73 & 4734 & K3-K5$^\dagger$ \\
    TOI-178 d   & 46   & 13.424  & 0.229 & $0.0095_{-0.0032}^{+0.0025}$ & 708.942  & Y & Y & 2 & False & 0.65 & 4316 & K7$^\dagger$ \\
    TOI-421 c   & 27   & 10.674  & 0.454 & $0.0517_{-0.0033}^{+0.0033}$ & 717.787  & Y & Y & 2 & False & 0.87 & 5325 & G7 \\
    WASP-29 b   & 29   & 9.770   & 0.770 & $0.2450_{-0.0220}^{+0.0230}$ & 941.816  & Y & Y & 2 & True & 0.79 & 4800 & K4$^\ddag$  \\ \hline

    GJ 9827 d   & 79   & 12.006  & 0.180 & $0.0127_{-0.0026}^{+0.0026}$ & 705.214  & Y & Y & 3 & False & 0.60 & 4340 & K6 \\
    HATS-72 b   & 57   & 21.946  & 0.722 & $0.1254_{-0.0039}^{+0.0039}$ & 714.490  & Y & Y & 3 & False & 0.72 & 4656 & K5$^\dagger$ \\
    HD 106315 c & 96   & 67.606  & 0.388 & $0.0478_{-0.0116}^{+0.0116}$ & 858.332  & N & Y & 3 & False & 1.30 & 6327 & F5 \\
    HD 63935 b  & 71   & 29.872  & 0.267 & $0.0340_{-0.0057}^{+0.0057}$ & 877.525  & Y & Y & 3 & False & 0.96 & 5534 & G5$^\ddag$ \\
    HD 63935 c  & 97   & 59.066  & 0.259 & $0.0349_{-0.0076}^{+0.0076}$ & 701.589  & N & Y & 3 & False & 0.96 & 5534 & G5$^\ddag$ \\
    HD 73583 b  & 54   & 14.314  & 0.249 & $0.0321_{-0.0098}^{+0.0107}$ & 688.149 & Y & Y & 3 & False & 0.65 & 4511 & K4 \\
    HD 97658 b  & 62   & 21.795  & 0.189 & $0.0261_{-0.0035}^{+0.0035}$ & 720.334  & Y & Y & 3 & True & 0.73 & 5212 & K1 \\
    K2-406 b    & 71   & 41.930  & 0.411 & $0.0603$ & 693.947  & N & Y & 3 & False & 0.96 & 5784 & G4 \\
    TOI-1130 b  & 83   & 21.998  & 0.326 & $0.0407$ & 786.160  & Y & Y & 3 & False & 0.69 & 4250 & K7 \\
    TOI-178 g   & 94   & 25.339  & 0.256 & $0.0124_{-0.0051}^{+0.0041}$ & 483.212  & N & Y & 3 & False & 0.65 & 4316 & K7$^\dagger$ \\
    TOI-620 b   & 61   & 9.190   & 0.335 & $0.0428$ & 620.967  & Y & Y & 3 & False & 0.55 & 3708 & M2.5 \\
    TOI-674 b   & 68   & 10.035  & 0.468 & $0.0743_{-0.0104}^{+0.0104}$ & 698.867  & Y & Y & 3 & False & 0.42 & 3514 & M2 \\
    V1298 Tau c & 83   & 48.488  & 0.499 & $0.0839$ & 932.639  & Y & Y & 3 & False & 1.34 & 4962 & K0 \\
    V1298 Tau d & 65   & 44.450  & 0.572 & $0.1060$ & 814.103  & Y & Y & 3 & False & 1.34 & 4962 & K0 \\
    WASP-11 b   & 57   & 17.901  & 1.110 & $0.5320_{-0.0200}^{+0.0210}$ & 918.615  & Y & Y & 3 & False & 0.89 & 4800 & K3 \\ \hline
    \end{tabular}
    \label{tab: Twinkle 4yr and 7yr candidates_Tier_1_2_3}
\end{table*}
\begin{table*}
    \caption{Candidate planets in Tiers $4$ and $5$ for the \textit{Twinkle} cool gaseous planet survey. Planetary and stellar parameters listed are used throughout this work and were obtained from the NASA Exoplanet Archive or calculated based on assumptions presented in \protect\cite{ARIELTargets_Edwards_2019}. Where unavailable in the archive, host star spectral type uses SIMBAD values ($^\ddag$) or estimation based on comparable stars ($^\dagger$).}
    \begin{tabular}{@{}l|rrrlr|cccc|ccc@{}}
    \toprule
      \textbf{Planet name} &
      \textbf{$\bm{N_\text{t}}$} &
      \textbf{$\bm{T_\text{int}}$} &
      \textbf{$\bm{R_\text{p}}$} & \textbf{$\bm{M_\text{p}}$} &
      \textbf{$\bm{T_\text{eq}}$} &
      \textbf{In 3yr} &
      \textbf{In 7yr} &
      \textbf{Tier} & 
      \textbf{Transmission} & \textbf{$\bm{R_\text{s}}$} & \textbf{$\bm{T_\text{s}}$} & \textbf{Spectral}\\ 
      & & \textbf{[days]} & \textbf{[\bm{$R_\text{J}$}]} & \textbf{[\bm{$M_\text{J}$}]} & \textbf{[K]} & \textbf{List} & \textbf{List} & & \textbf{Spectra} & \textbf{[$\bm{R_\odot}$]}& \textbf{[K]} & \textbf{Type}\\ \hline \hline
    HD 73583 c  & 136  & 60.023  & 0.213 & $0.0305_{-0.0054}^{+0.0057}$ & 510.880  & N & Y & 4 & False & 0.65 & 4511 & K4 \\
    K2-287 b    & 143  & 80.315  & 0.847 & $0.3150_{-0.0270}^{+0.0270}$ & 790.363  & N & Y & 4 & False & 1.07 & 5695 & G8 \\
    K2-32 b     & 126  & 54.426  & 0.473 & $0.0472_{-0.0054}^{+0.0057}$ & 808.226  & N & Y & 4 & False & 0.86 & 5271 & G9 \\
    LP 714-47 b & 143  & 27.265  & 0.419 & $0.0969_{-0.0047}^{+0.0047}$ & 686.753  & Y & Y & 4 & False & 0.58 & 3950 & M0 \\
    TOI-561 c   & 142  & 66.441  & 0.257 & $0.0220_{-0.0072}^{+0.0072}$ & 789.277  & N & Y & 4 & False & 0.85 & 5455 & G9 \\
    TOI-776 b   & 140  & 42.583  & 0.165 & $0.0126_{-0.0028}^{+0.0028}$ & 530.605  & N & Y & 4 & False & 0.54 & 3709 & M1 \\
    WASP-132 b  & 134  & 53.011  & 0.897 & $0.4100_{-0.0300}^{+0.0300}$ & 736.759  & Y & Y & 4 & False & 0.75 & 4714 & K4 \\
    WASP-84 b   & 123  & 42.392  & 0.942 & $0.6940_{-0.0470}^{+0.0490}$ & 773.076  & Y & Y & 4 & False & 0.75 & 5314 & K0 \\ \hline
    
    G 9-40 b    & 375  & 58.498  & 0.181 & $0.0150$ & 454.032  & N & Y & 5 & False & 0.31 & 3348 & M2.5 \\
    K2-121 b    & 309  & 96.569  & 0.671 & $0.1390$ & 786.419  & N & Y & 5 & False & 0.67 & 4690 & K5$^\ddag$ \\
    K2-138 f    & 181  & 72.914  & 0.259 & $0.0051_{-0.0037}^{+0.0067}$ & 715.911  & N & Y & 5 & False & 0.86 & 5356 & G8 \\
    LTT 3780 c  & 174  & 38.081  & 0.204 & $0.0198_{-0.0019}^{+0.0020}$ & 363.418  & N & Y & 5 & False & 0.37 & 3331 & M4 \\
    TOI-1201 b  & 234  & 44.232  & 0.215 & $0.0198_{-0.0028}^{+0.0026}$ & 682.774  & Y & Y & 5 & False & 0.51 & 3476 & M2 \\
    TOI-1422 b  & 173  & 96.480  & 0.353 & $0.0283_{-0.0063}^{+0.0072}$ & 838.973  & N & Y & 5 & False & 1.02 & 5840 & G2 \\
    TOI-1478 b  & 208  & 108.544 & 1.060 & $0.8510_{-0.0470}^{+0.0520}$ & 889.609  & N & Y & 5 & False & 1.05 & 5597 & G8 \\
    TOI-1634 b  & 1888 & 244.729 & 0.160 & $0.0319_{-0.0030}^{+0.0030}$ & 894.156  & N & Y & 5 & False & 0.45 & 3550 & M2 \\
    TOI-178 e   & 231  & 71.965  & 0.197 & $0.0121_{-0.0030}^{+0.0039}$ & 616.710  & N & Y & 5 & False & 0.65 & 4316 & K7$^\dagger$ \\
    TOI-3714 b  & 693  & 150.617 & 1.010 & $0.7000_{-0.0300}^{+0.0300}$ & 749.785  & N & Y & 5 & False & 0.51 & 3660 & M2 \\
    TOI-421 b   & 322  & 50.217  & 0.239 & $0.0226_{-0.0021}^{+0.0021}$ & 982.024  & N & Y & 5 & False & 0.87 & 5325 & G9 \\
    WASP-156 b  & 280  & 84.476  & 0.510 & $0.1280_{-0.0090}^{+0.0100}$ & 938.278  & Y & Y & 5 & False & 0.76 & 4910 & K3 \\ 
    WASP-8 b    & 236  & 94.124  & 1.130 & $2.1320_{-0.0810}^{+0.0800}$ & 896.199  & N & Y & 5 & False & 1.03 & 5600 & G8 \\ 
    Wolf 503 b  & 271  & 69.390  & 0.182 & $0.0197_{-0.0022}^{+0.0022}$ & 764.360  & N & Y & 5 & False & 0.69 & 4716 & K3.5 \\ \hline
 \end{tabular}
    \label{tab: Twinkle 4yr and 7yr candidates_Tier4_5}
\end{table*}

\bsp	
\label{lastpage}
\end{document}